\documentclass[12pt]{article}
\pdfoutput=1
\usepackage{jcapmod}
\usepackage{booktabs}
\usepackage{mathtools}
\usepackage[english]{babel}
\usepackage{amsmath,amssymb,amsbsy,amstext, amsthm}
\usepackage[usenames,dvipsnames]{xcolor}
\usepackage{hyperref}
\hypersetup{
    urlcolor=NavyBlue,
    citecolor=NavyBlue,
    linkcolor=NavyBlue,
}%
\usepackage{array}
\usepackage{url} 
\usepackage{slashed}
\usepackage{multicol}
\usepackage{blkarray}
\usepackage{enumerate}
\usepackage{graphicx}
\usepackage{amsfonts}
\usepackage{amssymb}
  \usepackage{empheq}
  \usepackage{tcolorbox}
\usepackage{xcolor}
\usepackage{colortbl}

\usepackage{float}
\usepackage[utf8]{inputenc}
\RequirePackage{color}
\usepackage[normalem]{ulem}
\usepackage{graphicx}

\usepackage{colortbl}
\definecolor{green2}{cmyk}{0, 1, 0.5, 0}
\definecolor{lightgreen}{cmyk}{0.2, 0, 0.2, 0.2}
\definecolor{lightgray}{cmyk}{0.1,0.2,0,0.1}
\definecolor{lightgray2}{cmyk}{0.4,0.4,0,0.8}
\definecolor{black}{cmyk}{1.0,1.0,1.0,1.0}

\allowdisplaybreaks[1]

\usepackage{colortbl}
\definecolor{lightgreen}{cmyk}{0.2, 0, 0.2, 0.2}
\definecolor{lightgray}{cmyk}{0.1,0.2,0,0.1}
\definecolor{lightgray2}{cmyk}{0.1,0.1,0,0.1}

\makeatletter
\newlength{\apb@width}
\newcommand{\autoparbox}[2][c]{\settowidth{\apb@width}{#2}\parbox[#1]{\apb@width}{#2}}

\makeatother

\numberwithin{equation}{section}

\def\be{\begin{equation}}
\def\ee{\end{equation}}

\def\bea{\begin{eqnarray}}
\def\eea{\end{eqnarray}}

\def\lp{\left(}
\def\rp{\right)}

\def\lb{\left[}
\def\rb{\right]}

\def\0{{\boldsymbol 0}}

\def\Gg{$\mathcal{G_{\gamma}}$}

\def\Sg{$\mathcal{S_\gamma}$}
\def\S{$\mathcal{S}$}
\def\F{$\mathcal{F}$}
\def\Kpm{$\mathcal{K}_\pm$}
\def\Kp{$\mathcal{K}_+$}
\def\km{$\mathcal{K}_-$}

\def\k{\mathbf{K}}
\def\w{\mathbf{W}}
\def\vp{\varphi}

\usepackage{setspace} 
\begin{document}

\begin{titlepage}

\setcounter{page}{1} \baselineskip=15.5pt \thispagestyle{empty}

\bigskip\

\vspace{1cm}
\begin{center}

{\fontsize{20}{28}\selectfont  \sffamily \bfseries { Extending the Dynamical Systems Toolkit: Coupled Fields in Multiscalar Dark Energy
 \\}}

\end{center}

\vspace{0.2cm}

\begin{center}
{\fontsize{13}{30}\selectfont Daniele Licciardello$^{a,b,c}$\textsuperscript{*}, 
Saba Rahimy$^{a,}$\textsuperscript{†},  Ivonne Zavala$^{a,}$\textsuperscript{‡}
} 
\end{center}

\begingroup
\renewcommand{\thefootnote}{\fnsymbol{footnote}}
\footnotetext[1]{\texttt{dani.licc21@gmail.com}}
\footnotetext[2]{\texttt{saba.rahimy@gmail.com}}
\footnotetext[3]{\texttt{e.i.zavalacarrasco@swansea.ac.uk}}
\endgroup

\begin{center}

\vskip 8pt
\textsl{$^a$ Centre for Quantum Fields and Gravity, Department of Physics, \\Swansea University, SA2 8PP, UK}\\
\textsl{$^b$ Dipartimento di Fisica e Astronomia, Universit\`a di Bologna,\\
  viale B. Pichat 6/2, 40127 Bologna,   Italy}\\
\textsl{$^c$ Dipartimento di Fisica e Astronomia Galileo Galilei , Universit\`a degli Studi di Padova, \\ \& I.N.F.N. Sezione di Padova, Via F. Marzolo 8, 35131 Padova, Italy
}
\vskip 6pt
\end{center}

\vspace{1.2cm}
\hrule \vspace{0.3cm}
\noindent
We study the dynamics of a two-field scalar model consisting of an axion-saxion pair with both kinetic and potential couplings, as motivated by string theory compactifications. We extend the dynamical systems (DS) toolkit by introducing a new set of variables that not only close the system and enable a systematic stability analysis, but also disentangle the role of the kinetic coupling.
Within this framework we derive a compact, general expression for the non-geodesicity (turning-rate) parameter evaluated at fixed points, valid for arbitrary couplings. This provides a transparent way of diagnosing non-geodesic dynamics, with direct applications to both dark energy and multifield inflation.
We first consider exponential coupling functions to establish analytic control and facilitate comparison with previous literature. In this case, we uncover a pair of genuinely non-geodesic fixed points, which act as  attractors within a submanifold of the full  system. In contrast, when the axion shift symmetry remains unbroken, our analysis shows that the apparent non-geodesic fixed point reported previously does not persist once the full dynamics are taken into account. 
Finally, we illustrate how our approach naturally extends to more realistic string-inspired models, such as power-law axion potentials combined with exponential saxion couplings, and present an explicit supergravity realisation.

\vskip 10pt
\hrule

\vspace{0.4cm}
 \end{titlepage}

 \tableofcontents

\section{Introduction}

Since the discovery of Hubble’s law, it has become clear that the Universe is expanding at an accelerated rate. The simplest way to incorporate this observation into general relativity is by introducing a positive cosmological constant in Einstein’s equations. However, observational data suggest that the required value of this constant is extremely small, implying a highly fine-tuned theory—something widely regarded as unnatural. Moreover, if the cosmological constant is indeed responsible for dark energy, we would be living in a de Sitter universe, which presents serious theoretical challenges for defining a consistent quantum field theory. In particular, as discussed in~\cite{Witten:2021jzq}, defining an S-matrix in a de Sitter background is problematic, making it difficult to formulate a well-behaved scattering theory.

From a quantum gravity perspective—especially within string theory—embedding a de Sitter solution has proven notoriously difficult. In such frameworks, the cosmological constant typically arises as the minimum of a scalar potential. Although various models have been proposed to achieve this, their robustness remains under debate. This situation led to the formulation of the de Sitter swampland conjecture, which asserts that consistent theories of quantum gravity may not allow for (meta)stable de Sitter vacua and places constraints on the slope of the potential even near positive-energy critical points~\cite{Garg:2018reu,Ooguri:2018wrx}.

On the observational side, there is growing evidence that dark energy may not be a true constant. For instance, recent results from the Dark Energy Spectroscopic Instrument (DESI)~\cite{DESI:2024kob,DESI:2024mwx,DESI:2025zgx,Lodha:2025qbg} and the Dark Energy Survey (DES)~\cite{DES:2024jxu,DES:2025bxy} suggest that models of dynamical dark energy—where the equation-of-state parameter $w_{\rm DE}$ varies with time—may offer a better fit to the data compared to the standard $\Lambda$CDM model with $w_{\rm DE} = -1$.

One of the most studied alternatives is quintessence~\cite{Ratra:1987rm,Peebles:1987ek,Caldwell:1997ii}, where dark energy is driven by a scalar field slowly rolling down a potential. This scenario, conceptually similar to inflation but occurring at late times, can generate accelerated expansion without invoking a cosmological constant. However, sustaining this behaviour requires the potential to be extremely flat. Constructing such slow-roll quintessence models in well-controlled regions of string moduli space appears to be unfeasible~\cite{Cicoli:2021fsd}, and remains a challenging task even in less controlled regimes~\cite{ValeixoBento:2020ujr,Cicoli:2021skd}.

A promising direction is to consider multi-field models, where more than one scalar field contributes to the dynamics. Such models can evade swampland constraints on the potential slope by allowing for strongly {\em non-geodesic} trajectories~\cite{Achucarro:2018vey,Chakraborty:2019dfh}, thereby achieving accelerated expansion even in steep potentials. These non-trivial trajectories generally require non-trivial couplings between the scalar fields—features that arise naturally in beyond-the-Standard-Model frameworks such as supergravity and string theory, which generically include a plethora of interacting scalar degrees of freedom.

In this context, many studies have focused on systems with two scalar fields, using dynamical systems techniques to investigate their cosmological evolution. In this approach, the equations of motion are reformulated into an autonomous system via a suitable set of dimensionless variables that typically describe the kinetic and potential energies of the fields.

Previous work has predominantly considered scenarios in which the scalar potential depends on only one field, often interpreted in string theory as the \emph{saxion}. The second field, typically an \emph{axion}, is assumed to preserve a continuous shift symmetry and thus does not appear in the potential. Axions descend from higher-dimensional $p$-form fields and acquire periodic or power-law potentials via symmetry breaking mechanisms, such as non-perturbative effects or background fluxes. Saxions, on the other hand, correspond to geometric moduli (e.g., the sizes and shapes of extra dimensions, or the dilaton) and commonly appear in the scalar potential exponentially. 

Cosmological solutions involving an axion-saxion system with a non-trivial field space metric (kinetic coupling) and a potential for the saxion have been studied recently in~\cite{Sonner:2006yn,Christodoulidis:2019jsx,Russo:2022pgo,Cicoli:2020cfj,Cicoli:2020noz,Akrami:2020zfz,Giacomini:2021kuf,Paliathanasis:2021fxi,Anguelova:2021jxu,Eskilt:2022zky,Brinkmann:2022oxy, Anguelova:2023dui,Revello:2023hro,Seo:2024qzf,Shiu:2024sbe}. In these works, the axion was treated as a free field, with no potential.
 A non-trivial potential for the axion was considered in~\cite{Marsh:2011gr,Marsh:2012nm}, although no kinetic coupling was included.
 Finally \cite{Paliathanasis:2020wjl,Paliathanasis:2023moe}, considered the special case in which the kinetic and potential couplings have the same functional form.
 Other recent works have explored multi-saxion models without kinetic coupling~\cite{Shiu:2023nph,Shiu:2023fhb,Seo:2024fki,Marconnet:2025vhj} (for a comprehensive review and earlier work see~\cite{Bahamonde:2017ize} and references therein).

Motivated by the structure of typical scalar sectors  in string theory  compactifications—where axions and saxions are generically coupled through both kinetic and potential terms—and by the prospects of realising accelerated expansion via non-geodesic dynamics, we extend previous studies by incorporating a non-trivial potential coupling for the axion-saxion system. We analyse an axion-saxion system with both kinetic and potential couplings using the framework of dynamical systems. To the best of our knowledge, this is the first time such a system has been successfully closed and systematically studied within this formalism. As we will demonstrate, achieving closure of the system and performing a coherent analysis of the stability requires a non-trivial extension of the standard approach. 

The paper is organised as follows. In Section~\ref{sec:ASS}, we introduce the axion--saxion system, outline the cosmological equations, and discuss the general structure of multifield dynamics. Section~\ref{sec:DS} develops the full dynamical system and presents the new variables that allow us to close it and disentangle the role of the kinetic coupling. Within this framework, we also derive a compact and general expression for the non-geodesicity (turning-rate) parameter at fixed points, valid for arbitrary couplings. We then analyse the case of purely exponential couplings, which permits analytic control and a direct comparison with existing literature. This analysis reveals the existence of a genuinely non-geodesic sub-attractor and clarifies the apparent non-geodesic behaviour in the flat axion case.  Finally, we illustrate how our approach can be applied to more realistic potentials, such as monomial axion potentials motivated by axion monodromy. In Section~\ref{sec:sugra}, we embed this construction in a string-inspired supergravity model, providing a concrete ultraviolet realisation. 
We conclude in Section~\ref{sec:conclusion} with a summary of our findings and an outlook toward future generalisations and applications. Some technical details are relegated to two final Appendices, \ref{app:SFP} and \ref{app1}.

\section{The Axion-Saxion System}\label{sec:ASS}

We are interested in multiscalar sectors arising from typical string theory compactifications, whose low-energy effective field theory is governed by ${\cal N}=1$ supergravity (sugra). In this framework, scalar degrees of freedom are organised into chiral multiplets, whose scalar component is a complex field. 
The real and imaginary parts of this field correspond respectively, to the saxion and axion fields that form the focus of our analysis.

Expressed in terms of real fields, the low-energy effective action takes the form\footnote{We set $M_{\rm Pl} = 1$ unless otherwise stated.}:
\be\label{eq:genaction}
S = \int d^4 x \, \sqrt{-g} \left[\frac{1}{2} R - \frac{1}{2} \gamma_{ab}(\phi) \, \partial_\mu \phi^a \partial^\mu \phi^b - V(\phi^a) \right]\,,
\ee
where $\gamma_{ab}$ denotes the {\em field space metric}. 
For the axion-saxion system under consideration, this metric can generically be written in the form\footnote{This metric can be expressed in various equivalent forms via simple field redefinitions:
\be
ds^2 = dR^2 + f^2(R)\, d\theta^2 = f^2(r)\left(dr^2 + d\theta^2\right) = f^2(\rho)\left(d\rho^2 + \rho^2 d\theta^2\right)\,.
\ee
}:
\be\label{eq:metric}
\gamma_{ab} = {\rm diag}\left(1, f^2(\phi)\right)\,,
\ee
where $\phi$ is the saxion, and $\chi$ is the axion. Due to the axionic shift symmetry, the metric is independent of $\chi$, and the function $f(\phi)$ encodes the {\em kinetic coupling} between the axion and saxion.

In addition to kinetic interactions, the two fields generically interact through the scalar potential, which we take to have the following form:
\be\label{eqn:shape V}
V(\phi, \chi) = W(\phi) + g(\phi)\, U(\chi)\,,
\ee
where the function $g(\phi)$ captures the {\em potential coupling} between the saxion and axion. The scalar sector of the theory can then be written explicitly as:
\be\label{eq:scalarL}
{\cal L}_s = -\frac{1}{2} \partial_\mu \phi \partial^\mu \phi - \frac{f^2(\phi)}{2} \partial_\mu \chi \partial^\mu \chi - W(\phi) - g(\phi)\, U(\chi)\,.
\ee
Finally, we note that the curvature of the field space metric is given by
\be\label{eq:Rfs}
R_{\rm fs} = -\frac{2 f_{\phi\phi}}{f}\,,
\ee
where $f_\phi \equiv \partial_\phi f$. For instance, a flat field space corresponds to $f_{\phi\phi}=0$, which implies that $f(\phi)$ is linear in $\phi$. By contrast, field spaces of constant curvature are characterised by $\tfrac{f_{\phi\phi}}{f} = \text{const.}$, which leads to an exponential form for $f(\phi)$.

\subsection{Cosmological evolution}

To study the late-time dynamics of the universe, we consider a spatially flat FLRW metric with line element:
\be
ds^2 = -dt^2 + a^2(t)\, dx_i dx^i\,.
\ee
We include a matter component modeled as a barotropic fluid with pressure $p = w\rho$. For most of our analysis, we keep $w$ general, although for late-time acceleration the case of interest is $w = 0$, corresponding to pressureless matter (i.e., dark matter and baryons). Radiation can – in principle – be included, but is negligible at late times and will thus be omitted from our analysis.
The evolution of the system is governed by the coupled Einstein-scalar equations:
\begin{subequations}\label{eq:fulleoms}
 \begin{align}
     &H^2 = \frac{1}{3}\rho_\varphi + \frac{1}{3}\rho \,, \label{eq:Friedmann}\\
     &\ddot\phi^a +3H\dot\phi^a +\Gamma^a_{\,\,bc}\dot\phi^b\dot\phi^c +\gamma^{ab}\partial_b V = 0\,,\label{eq:ScalarsEq}
 \end{align}
\end{subequations}
where $\Gamma^a_{\,\,bc}$ are the Christoffel symbols associated with the scalar field space metric $\gamma_{ab}$, and $\partial_b V$ denotes the derivative of the potential with respect to $\phi^b$.

The scalar sector contributes to the total energy density and pressure via:
\be
\rho_\varphi = \frac{1}{2} \dot\varphi^2 + V(\phi^a) \,, \qquad 
p_\varphi = \frac{1}{2} \dot\varphi^2 - V(\phi^a)\,,
\ee
with the total scalar kinetic energy defined by
\be
\dot\varphi^2 = \gamma_{ab} \dot\phi^a \dot\phi^b\,.
\ee

Using the explicit Lagrangian \eqref{eq:scalarL}, the equations of motion for the saxion $\phi$ and axion $\chi$ become\footnote{This two-field system can alternatively be written in terms of a dynamical vector $Q^\nu$ as shown in \cite{Rahimy:2025iyj}.}:
\begin{subequations}\label{eq:eomscalars}
     \begin{align}
         \ddot \phi + 3H\dot\phi - f f_\phi\,\dot \chi^2 +  W_\phi + g_\phi \, U &= 0 \,,\\
         \ddot \chi + 3H\dot\chi + 2\frac{f_\phi}{f} \dot\phi\, \dot\chi + \frac{g \, U_\chi}{f^2} &= 0\,,
     \end{align}
\end{subequations}
where we used the explicit potential \eqref{eqn:shape V} and $f_\phi$, etc. denote  derivative of $f$ with respect to $\phi$, etc.  

\subsection{Multifield dynamics}\label{sec:multidyn}

In the presence of multiple scalar fields, the dynamics of the system becomes significantly richer than in the single-field case. The evolution of the scalars in field space can exhibit highly non-trivial behaviour, which is most naturally characterised by the degree to which their trajectory deviates from geodesic motion. This deviation is quantified by the so-called \emph{turning rate} parameter, which measures the curvature of the path in field space. Throughout this work, we adopt the terminology \emph{geodesic} (${\cal G}$) and \emph{non-geodesic} (${\cal NG}$) to distinguish between trajectories with vanishing and non-vanishing turning rates, respectively.\footnote{In the inflationary literature, the term \emph{turning rate} is more commonly used to describe this phenomenon.}

We now introduce the (dimensionless) non-geodesicity parameter, $\boldsymbol{\omega}$, which plays a central role in identifying genuine multifield trajectories—namely those that strongly deviate from geodesic motion.
To define this quantity, we employ a kinematic basis to decompose the scalar trajectory into tangent and normal directions in field space \cite{Sasaki:1995aw,Gordon:2000hv,GrootNibbelink:2001qt}. This is achieved by introducing the unit tangent and normal vectors, $T^a$ and $N^a$, defined as
\begin{equation}
T^a \equiv \frac{\dot\phi^a}{\dot\varphi}\,, \qquad 
N_a \equiv \sqrt{\det{\gamma_{ab}}}\,\epsilon_{ab}\,T^b\,,
\end{equation}
which satisfy the orthonormality relations
\[
T^a T_a = 1\,, \qquad
N^a T_a = 0\,, \qquad
N^a N_a = 1\,.
\]

The scalar equations of motion \eqref{eq:ScalarsEq} can then be projected along these two directions, yielding
\begin{align}
\ddot\varphi + 3H\dot\varphi + V_T &= 0\,, \label{varphiT} \\
D_t T^a + \boldsymbol{\Omega} N^a &= 0\,, \label{varphiN}
\end{align}
where $V_T \equiv V_a T^a$ and $V_N \equiv V_a N^a$ are the tangential and normal projections of the potential gradient, respectively. The parameter $\boldsymbol{\Omega}/H$ quantifies the deviation from geodesic motion and is given by
\begin{equation}\label{Omega}
\boldsymbol{\omega} \equiv \frac{\boldsymbol{\Omega}}{H} \equiv \frac{V_N}{H\dot\varphi}\,.
\end{equation}
Here, $D_t T^a$ denotes the covariant time derivative in field space:
\begin{equation}\label{Dt}
D_t T^a \equiv \dot T^a + \Gamma^{a}_{bc} T^b \dot \phi^c\,.
\end{equation}

Note that the normal component of the potential gradient satisfies
\[
V_N N^a = V^a - V_T T^a\,,
\]
and geodesic motion corresponds to the case where $V^a = V_T T^a$, that is, when the potential gradient is fully aligned with the tangent vector.

For the axion-saxion system defined in \eqref{eq:scalarL}, the tangent and normal vectors take the explicit form
\begin{equation}
T^a = \frac{1}{\dot\varphi}\left(\dot\phi,\dot\chi\right)\,, \qquad 
N^a = \frac{f}{\dot\varphi}\left(\dot \chi, -\frac{\dot\phi}{f^2}\right)\,.
\end{equation}
Consequently, the non-geodesicity parameter $\boldsymbol{\omega}$ becomes
\begin{equation}\label{eq:NGparam}
\boldsymbol{\omega} = \frac{1}{H f \dot\varphi^2} \left( f^2 \dot\chi\, V_\phi - \dot\phi\, V_\chi \right)\,.
\end{equation}
Genuine multifield trajectories are characterised by $\boldsymbol{\omega} \gtrsim 1$.

\section{Dynamical systems analysis}\label{sec:DS}

 We are now ready to analyse the cosmological evolution of the axion–saxion system in the presence of an additional barotropic fluid using dynamical systems techniques. We refer the reader to \cite{Bahamonde:2017ize} for  a  recent review on dynamical systems techniques in cosmology for details on this approach. 

Let us first pause to summarise previous related work.
As mentioned in the introduction, axion–saxion cosmology has been previously studied using these techniques, particularly in scenarios where the axion shift symmetry remains unbroken, that is, where $U_\chi = 0$ in our notation \cite{Sonner:2006yn,Christodoulidis:2019jsx,Russo:2022pgo,Cicoli:2020cfj,Cicoli:2020noz,Akrami:2020zfz,Anguelova:2021jxu,Eskilt:2022zky,Brinkmann:2022oxy,Anguelova:2023dui,Revello:2023hro,Seo:2024qzf,Shiu:2024sbe}. However, in general, the axion acquires a potential—either from non-perturbative effects or, in some even at tree level (axion monodromy)—and thus cannot be treated as a flat direction.

On the other hand, \cite{Marsh:2011gr,Marsh:2012nm} incorporated an axion potential, but did not consider any kinetic coupling between the axion and saxion (i.e., $f_\phi = 0$). In contrast, in supergravity and string-derived effective theories, both types of couplings—kinetic and potential—are generically present. Their combined presence significantly enriches the structure of the cosmological evolution, especially when multifield effects are taken into account. Indeed, genuine multifield trajectories—those with strong deviations from geodesics in field space—open up qualitatively new dynamical possibilities for both early- and late-time acceleration.

Including both kinetic and potential couplings poses a nontrivial challenge for the standard dynamical systems approach to scalar cosmology. In particular, it complicates the task of constructing a closed autonomous system: that is, a system of first-order differential equations whose evolution depends only on a finite set of (dimensionless) variables and not on external functions or higher-order derivatives.
We address this by identifying a new set of dynamical variables tailored to the axion-saxion set-up that allow the system to close, even in the presence of nontrivial field-space curvature and axion–saxion potential couplings. With these variables in hand, we will investigate the fixed points of the system, analyse their stability properties, and characterise the full cosmological dynamics. 
We also provide a neat expression for the non-geodecisity parameter at the fixed points of the general system. 
We will then compare our findings to those in the existing literature and highlight the new qualitative behaviours  that arise when both kinetic and potential couplings are treated simultaneously.

\subsection{Dynamical variables}
  
As  mentioned above, the kinetic coupling, $f$, represents a challenge and  a natural variable choice would be to use $f$ (or some function of it) directly as a variable. However, one can check that this choice is not suitable to extract the full set of fixed points that can appear, and to  make  the connection to previous results in the case of a flat axion. In appendix \ref{app1} we show an alternative set of variables and outline the problems that arise in that case. 

Through a non-trivial process of exploration and refinement, we arrive at the following minimal set of variables, which captures the full dynamics of the system and enables a thorough stability analysis, as well as to close the system: 
    \be\label{eq:Dvariables1}
        x_1=\frac{\dot\chi}{\sqrt{6} H}\,, \quad 
        x_f= \frac{f \dot\chi}{\sqrt{6} H}\,, \quad 
        x_2=\frac{\dot\phi}{\sqrt{6} H}\,, \quad
        y_1=\frac{\sqrt{g(\phi)\,U(\chi)}}{\sqrt{3}H}\,,   \quad y_2=\frac{\sqrt{W(\phi)}}{\sqrt{3}H} \,, 
    \ee
    \be\label{eq:Dvariables2}
        y_f= \frac{y_1}{f} \, , \quad \Omega = \frac{\rho}{3 H^2}  \, , \quad \lambda_1=-\frac{U_\chi}{U}\,, \quad \lambda_2=-\frac{W_\phi}{V}\,,\quad \gamma= -\frac{g_\phi}{g} \, , \quad \beta =\frac{f_\phi}{f}\,.  
    \ee
    With these definitions the Friedman constrain becomes:
\be\label{eq:FriedConstr}
        1 = x_f^2 + x_2^2 + y_1^2+ y_2^2+ \Omega\,,
    \ee
    and we rewrite it as:
    $\Omega = 1- x_f^2 - x_2^2 - y_1^2- y_2^2$. 
In terms of these variables, the equations of motion  \eqref{eq:fulleoms} can be written as\footnote{Interestingly, the equation for $y_f$ features a factor involving the potential and kinetic couplings of the saxion to the axion, appearing through the combination $\gamma+2\beta$. This same structure has emerged in previous studies of coupled scalar dark sectors in \cite{Rahimy:2025iyj}.}
    \begin{subequations}\label{eq:eomsDS}
        \begin{align}
            x_1' &= 3 \, x_1 ( x_f^2 + x_2^2 - 1 ) - 2 \sqrt{6} \beta x_1 x_2 + \sqrt{\frac{3}{2}} \lambda_1 \,y_f^2  + \frac{3}{2} x_1 (1+w)\Omega \,\label{eq:eomsx1} ;\\
            x_f' & = 3 \, x_f ( x_f^2 + x_2^2 - 1 ) - \sqrt{6} \beta x_f \,x_2 + \sqrt{\frac{3}{2}} \lambda_1 \,y_f \,y_1  + \frac{3}{2} x_f (1+w)\Omega \,\label{eq:eomsxc} ; \\
            x_2' & = 3 \, x_2 ( x_f^2 + x_2^2 - 1 ) + \sqrt{6} \beta x_f^2 + \sqrt{\frac{3}{2}} \lp \gamma \,y_1^2 + \lambda_2\, y_2^2 \rp  + \frac{3}{2} x_2 (1+w)\Omega \,\label{eq:eomsx2} ; \\
            y_1' & = \frac{\sqrt{6}}{2}  y_1 \, \lb - \lambda_1 x_1 - \gamma x_2 + \sqrt{6} (x_f^2 + x_2^2) +\sqrt{\frac{3}{2}}  (1+w)\Omega \rb \, \label{eq:eomsy1}; \\
            y_2' & = \frac{\sqrt{6}}{2}  y_2 \, \lb - \lambda_2 x_2 + \sqrt{6} (x_f^2 + x_2^2) +\sqrt{\frac{3}{2}}  (1+w) \Omega \rb \, \label{eq:eomsy2}; \\
            y_f' & = \frac{\sqrt{6}}{2}  y_f \, \lb - \lambda_1 x_1 - (\gamma + 2 \beta) x_2 + \sqrt{6} (x_f^2 + x_2^2) +\sqrt{\frac{3}{2}}  (1+w) \Omega \rb \, \label{eq:eomsyk}; \\
            \beta' &= \sqrt{6} \left[ \Gamma_f - 1 \right] \beta^2 \,x_2 \, \label{eq:betap}; \\
            \gamma' &= -\sqrt{6} \lb \Gamma_g -1 \rb \gamma^2 \,x_2 \, ;  \\
            \lambda'_1 &= -\sqrt{6} \lb \Gamma_1 -1 \rb \lambda_1^2 \,x_1 \, ; \label{eqn:eom_lambda1} \\
            \lambda_2' &= -\sqrt{6} \lb \Gamma_2 -1 \rb \lambda_2^2\, x_2 \, \label{eq:lam2p} ; 
        \end{align}
    \end{subequations}
    where we defined:
    \be \label{eqn: Gammas def}
        \Gamma_f = \frac{f_{\phi \phi} f }{(f_\phi)^2} \, , \quad \Gamma_1 = \frac{U_{\chi \chi} U }{(U_\chi)^2} \, , \quad \Gamma_g = \frac{g_{\phi \phi} g }{(g_\phi)^2} \, , \quad \Gamma_2 = \frac{W_{\phi \phi} W }{(W_\phi)^2} \,.
    \ee

We can further derive expressions for the equations of state, both for the scalar sector, $w_\varphi$, and for the full system, $w_{\rm eff}$, relevant for the cosmological evolution,  in terms of the variables above as follows:
\bea
w_\varphi = \frac{  x_f^2 + x_2^2 - y_1^2-y_2^2}{ x_f^2 + x_2^2 + y_1^2 + y_2^2} \,,\qquad \\
w_{\rm eff} =   x_f^2 + x_2^2 - y_1^2 - y_2^2 + w \,\Omega\,,
\eea
where $w_\varphi =p_\varphi/\rho_\varphi$, and $w_{\rm eff}= p_{\rm eff}/\rho_{\rm eff}$, with $p_{\rm eff}= p_\varphi+p$. In particular, an accelerated expansion of the system is achieved for  $w_{\rm eff}<-1/3$, which can be seen from the requirement that the acceleration parameter $\epsilon<1$, where 
\be
\epsilon\equiv -\frac{\dot H}{H^2} = \frac{3}{2}(1+w_{\rm eff})\,.
\ee
    Notice that for  general potentials, the quantities in \eqref{eqn: Gammas def} are not constant, and the system will not be closed. However in several instances of potentials, these do indeed become constant and the system closes. 

The physical phase space can be restricted by considering the Friedmann constraint \eqref{eq:FriedConstr}, which implies that $0\leq \Omega \leq1$, as well as $0\leq \Omega_\varphi \leq 1$, where 
\be
\Omega_\varphi \equiv \frac{\rho_\varphi}{3H^2} = x_f^2  +x_2^2 +y_1^2+y_2^2\,,
\ee
which restricts $0\leq x_f^2\leq1$, $0\leq x_2^2\leq1$, $0\leq y_1^2\leq1$, $0\leq y_2^2\leq1$.
Moreover, from the equations \eqref{eq:eomsDS} we see that $y_1=0, y_2=0$ and $y_f=0$ are invariant sets (or submanifolds) of the system\footnote{An invariant set of the system is a region of the phase-space where one of the coordinates  is constant, with a value which satisfies its own equation of motion independently of the value of the others. This concept can be extended to more the one variables in a simple way.}. In the case of constant $\beta, \gamma, \lambda_{1,2}$, the system is further invariant under the following symmetries
    \begin{subequations} \label{eqn:syms}
    \begin{align}
        x_1 & \rightarrow - x_1  , x_f \rightarrow - x_f, \lambda_1 \rightarrow - \lambda_1\,, \label{eqn:sym1} \\
        x_2 &\rightarrow - x_2 , \lambda_2 \rightarrow - \lambda_2 , \beta \rightarrow - \beta ,\gamma \rightarrow - \gamma \,,\label{eqn:sym2} \\
        x_f & \rightarrow - x_f\,,\, 
        y_f \rightarrow - y_f\\
        y_1 & \rightarrow - y_1\, , \, y_f \rightarrow - y_f \label{eqn:sym3}\,, \\
        y_2 & \rightarrow - y_2 \label{eqn:sym4}\,.
    \end{align}
    \end{subequations}
Therefore in this case, one can focus on positive values of some of the variables and parameters, specifically,  $y_{1,2}, \lambda_{1,2}$.

\subsubsection{The non-geodesicity parameter, $\boldsymbol{\omega}$}

As discussed before, once more than one scalar field is present, the dynamics become considerably richer. In particular, new phenomena arise due to the genuine multifield nature of the system. A useful way to characterise this is through the non-geodesicity (or turning rate) parameter, $\boldsymbol{\omega}$, introduced in Section~\ref{sec:multidyn} (see eq.~\eqref{eq:NGparam}).  

In terms of the dynamical variables \eqref{eq:Dvariables1}, this takes the form
\be\label{eq:NGw}
\boldsymbol{\omega} = \sqrt{\tfrac32}\,\frac{1}{\lp x_f^2+x_2^2\rp}\Big( \lambda_1 \,x_2\,y_1\,y_f \;-\; x_f\big[\gamma \,y_1^2 + \lambda_2\,y_2^2\big]\Big)\,.
\ee
From this expression we can clearly distinguish the kinetic and potential contributions to the turning rate.  

Our focus will be on the critical (fixed) points of the system \eqref{eq:Dvariables1}, defined by ${\bf x}'=0$, where ${\bf x}$ denotes the vector of dynamical variables. At such points the system is at rest, and remarkably, the turning rate simplifies to  
\be\label{eq:wc}
\boldsymbol{\omega}_c = \sqrt{6}\,\beta \,x_{f,c}\,.
\ee
That is, at critical points, $\boldsymbol{\omega}$ depends solely on $x_f$ and  the fractional rate of change of the field space metric, encoded in the ratio $f_\phi/f=\beta$.  This has  important consequences:  

\begin{enumerate}[i)]

  \item If $\beta=0$ (i.e.~$f=\text{const.}$), the fixed points are necessarily geodesic.  

  \item  Non-geodesic fixed points arise  when $x_{f,c}\neq 0$, though their physical relevance must be checked against reality and consistency conditions as we will see later. 
  
\item Non-geodesic fixed points are independent of the saxion kinetic energy, $x_2$, which can therefore take any value at the fixed point. Similarly, they are  in principle independent of the potential energies $y_i$, however, for specific potentials, the physical relevance must be checked against reality and consistency conditions.

\item  Since   $x_f^2\leq 1$,   the magnitude of the turning rate depends directly on the  fractional rate of change of the field space metric, $\beta$.
  
\end{enumerate}

\subsubsection{The variables $x_1, x_f$, and the kinetic coupling}

Let us now pause to outline the rationale behind our choice of dynamical variables. In particular, due to the presence of a kinetic coupling, it becomes necessary to introduce the variable \( x_1 \) in order to close the system. As defined in \eqref{eq:Dvariables1}, \( x_1 \) plays a crucial role in disentangling the effects of the kinetic coupling function \( f(\phi) \) from the intrinsic velocity of the axion field, \( \dot{\chi} \). Together with \( x_f \), which encodes the axion's kinetic energy, the variable \( x_1 \) allows us to clearly distinguish between different dynamical regimes involving the coupling.

When analysing the fixed points of the system, the values of \( x_1 \) and \( x_f \) reveal important information about the behavior of the coupling \( f \) and the axion dynamics. In particular, the following cases can be identified:
\begin{itemize}
        \item \( x_1 = 0 \) and \( x_f = 0 \): this corresponds to a static axion (\( \dot{\chi} = 0 \)), and therefore vanishing axion kinetic energy.
    
    \item \( x_1 = 0 \) and \( x_f \ne 0 \): this implies \( f \to \pm \infty \), since the axion velocity vanishes at the fixed point. In order for the kinetic energy (which depends on the product \( f\,\dot\chi \)), to remain finite and non-zero, the coupling \( f \) must diverge so as to compensate for the vanishing velocity.

    \item \( x_1 \ne 0 \) and \( x_f = 0 \): in this case, the axion is moving (\( \dot{\chi} \ne 0 \)), but its kinetic energy vanishes. This can only happen if the kinetic coupling vanishes, i.e., \( f = 0 \).
    
    \item \( x_1 \ne 0 \) and \( x_f \ne 0 \): this is the general case, corresponding to a non-vanishing axion velocity and a finite kinetic coupling, \( f \ne 0 \).
 
\end{itemize}

This classification will be useful in interpreting the physical nature of the fixed points and their corresponding trajectories in phase space.

\subsubsection{The need for the variable $y_f$}\label{subsec:yf}
    As we have seen, the introduction of the variable \( x_1 \) allows us to close the system of equations. However, for the system to be well-defined and analytically tractable--particularly near the region where \( x_f \to 0 \) --we find it necessary to introduce an additional variable, defined in \eqref{eq:Dvariables2}.
To understand the motivation for this choice, let us consider what happens if we attempt to work with the original variables alone. In the evolution equations for \( x_1' \) and \( x_f' \), one encounters terms involving ratios  \( \left(\frac{x_1}{x_f}\right)^n\), which become singular as \( x_f \to 0 \). Following the standard method described in \cite{Bahamonde:2017ize}, one way to circumvent such singularities is to multiply the entire system of equations by \( x_f^2 \) -- the highest power of \( x_f \) appearing in any denominator. This regularises the equations and avoids explicit divergences.

However, this procedure comes at a cost: by multiplying through by $x_f^2$, the system of equations loses all its linear terms. As a result, the Jacobian matrix evaluated at any fixed point becomes identically zero, making linear stability analysis impossible. One is then forced to perform a second-order (or higher) analysis of the dynamical system, which is cumbersome and intractable given the dimensionality of the system. 

This is precisely where the new variable \( y_f \) becomes useful. The problematic terms in the equations typically originate from the axion's equation of motion, where the kinetic coupling appears in the denominator. Fortunately, such terms are always accompanied by a factor of \( y_1^2 \), allowing us to absorb the kinetic function \( f \) into a redefinition. By defining \( y_f = y_1 / f \), we avoid introducing singularities when \( x_f \to 0 \), while simultaneously preserving the linear structure of the system. This makes the analysis of fixed points and their stability significantly more manageable.   

Finally, note that the variables \( x_1, x_f \) do not enter the Friedmann constraint or any other direct physical constraint equations. As a result, their values are not bounded by those relations and thus they take values in the range $x_1\in(-\infty,\infty), \,\, y_f\in (-\infty,\infty)$.

\subsection{Exponential potentials and couplings}

We start by considering  the case 
 where the functions \( f(\phi), U(\phi), g(\phi), W(\phi) \) are exponentials. This makes the system analytically tractable and will allow us to make  a comparison with earlier studies.
 For this case,  the associated parameters \( (\beta, \gamma, \lambda_1, \lambda_2) \) are all  constants making their associated equations \eqref{eq:betap}–\eqref{eq:lam2p}  trivially satisfied, and reducing the  dynamical system to  a six-dimensional autonomous system. 

This choice covers a wide class of models studied in the literature in the flat axion case, and provides a useful framework to identify and classify fixed points. We now proceed to analyse these fixed points and study their physical properties.

\subsubsection{Fixed points and their properties}\label{subsec:FPs}
    The fixed points for the axion-saxion system \eqref{eq:eomsDS} are summarised in Table \ref{tab:FPs_Exp}. 
Because of the additional variables required to close our dynamical system, the fixed-point analysis here departs slightly from the standard scalar-field cosmology treatment. In particular, the existence conditions are not determined solely by the usual algebraic constraints but also by relations inherited from the underlying functional dependence of the coupling and potential terms on the scalar fields.
Below we analyse each fixed point in detail and explain how its existence conditions are obtained.
In addition to the usual requirements -- namely, the reality of all dynamical variables and the Friedmann constraint \eqref{eq:FriedConstr} -- we must also impose further consistency conditions arising from the functional dependence  on the scalar fields\footnote{Note that in dynamical-systems approach to scalar cosmologies, some fixed points correspond to asymptotic regimes in which the original scalar fields diverge. This simply reflects that the phase-space variables remain finite while the fields approach limiting regions of the potential, and such solutions are reached only asymptotically in cosmic time.
}. We spell out these extra restrictions explicitly in what follows. The cosmological parameters for each point are listed in Table \ref{tab:CosmoParams_Exp}.

\bigskip

\begin{table}[h]
    \hskip-1.2cm\resizebox{18.5cm}{!}{
    \begin{tabular}{|c|c|c|}
        \hline
        \rowcolor{gray!30}
        \textbf{Point} & $(x_1,x_f, x_2, y_1, y_2,y_f)$  & \textbf{Existence conditions} \\ 
        \hline
                \hline
        \Kpm & $\lp 0,0,\pm 1,0,0,0 \rp$ & $\lambda_2 > 0 \bigwedge \beta \leq 0 \bigwedge $  \\
        &  & $  \lb \lp \gamma + 2\beta \geq 0 \bigwedge \lambda_1 > 0 \rp \bigvee \lp \gamma + 2\beta > 0 \bigwedge \lambda_1 \geq 0 \bigwedge \gamma \neq 0 \rp \rb$   \\
        \hline
       $\widetilde{\mathcal K}_- $& $\lp \frac{\sqrt{6} \, + 2\beta + \gamma}{\lambda_1}, 0, -1, 0, 0, \pm \frac{2 \sqrt{-\beta \, (\sqrt{6} \, + 2\beta + \gamma)}}{\lambda_1} \rp$ & $\lambda_1>0 \bigwedge \beta<0 \bigwedge \lb \gamma,\lambda_2>0 \bigvee \lp\gamma>0 \bigwedge \lambda_2=0 \rp \bigvee \lp \gamma=0 \bigwedge \lambda_2 > 0\rp \rb \bigwedge $  \\
       &  & $  \gamma > - \, \sqrt{6} \, - 2\beta $  \\
        \hline
      $\widetilde{\mathcal K}_+$ & $\lp \frac{\sqrt{6} \, - 2\beta - \gamma}{\lambda_1}, 0, +1, 0, 0, \pm \frac{2 \sqrt{\beta \, (\sqrt{6} \, - 2\beta - \gamma)}}{\lambda_1} \rp$ & $\lambda_1 > 0 \bigwedge \lambda_2\geq 0 \bigwedge \beta<0 \bigwedge \gamma > \sqrt{6} - 2 \beta$ \\
        \hline
        \F & $(0,0,0,0,0,0)$ & $ \lambda_2 > 0 \bigwedge \beta \leq 0 \bigwedge $ \\
        & & $  \lb \lp \gamma + 2\beta \geq 0 \bigwedge \lambda_1 > 0 \rp \bigvee \lp \gamma + 2\beta > 0 \bigwedge \lambda_1 \geq 0  \bigwedge \gamma \neq 0 \rp \rb$ 
        \\
        \hline
        $\widetilde{\mathcal F}_\pm$ & $\lp \sqrt{\frac{3}{2}} \frac{1+ w}{\lambda_1} ,0,0,0,0, \pm \sqrt{\frac{3}{2}} \frac{\sqrt{1- w^2}}{\lambda_1} \rp$ & $\lambda_1, \lambda_2>0 \bigwedge \beta<0 \bigwedge \gamma \geq0$ \\ 
        \hline
        ${\mathcal S}_{\lambda_2}$  & $\lp 0, 0, \sqrt{\frac{3}{2}} \frac{1+ w}{\lambda_2}, 0, \sqrt{\frac{3}{2}} \frac{\sqrt{1-w^2}}{\lambda_2}, 0 \rp$ & $\lambda_1 \geq0 \bigwedge \lambda_2 \geq \sqrt{3(1+w)};\,\,\,\beta\in \mathbb{R} $  \\
        \hline
        \Sg & $\lp 0, 0, \sqrt{\frac{3}{2}} \frac{1+w}{\gamma}, \sqrt{\frac{3(1-w^2)}{2\gamma^2}},0,0 \rp$ & $\lambda_1, \lambda_2,\beta >0 \bigwedge \lp \gamma \leq - \sqrt{3 (1+w)} \, \bigvee \gamma \geq \sqrt{3 (1+w)} \rp$ \\
        \hline
        ${\mathcal G}_{\lambda_2}$  & $\lp 0, 0, \frac{\lambda_2}{\sqrt{6}}, 0, \sqrt{1- \frac{\lambda_2^2}{6}}, 0 \rp$ & $\lambda_1 \geq 0 \bigwedge \lambda_2 \in [ 0, \sqrt{6} );\,\,\,\beta,\gamma\in \mathbb{R} $  \\
        \hline
        \Gg &  $\lp 0, 0, \frac{\gamma}{\sqrt{6}}, \sqrt{1-\frac{\gamma^2}{6}}, 0, 0 \rp$ & $ \lambda_2,\beta > 0 \bigwedge \lambda_1 \geq 0 \bigwedge \gamma \in ( - \sqrt{6} \, , \sqrt{6}) \, $\\
        \hline
        $\mathcal{G_{\chi\pm}}$ or $\mathcal{K_{\chi\pm}}$ & $\lp \frac{\sqrt{6} \gamma}{\lambda_1 (\gamma - 2 \beta)}, \pm\frac{\sqrt{\gamma}}{\sqrt{\gamma - 2 \beta}}, 0, \sqrt{\frac{2 \beta}{2 \beta - \gamma}}, 0, \frac{2}{\lambda_1} \sqrt{ - \frac{3 \beta \gamma}{(2 \beta - \gamma)^2}} \rp$ & 
        $\lambda_1, \lambda_2 >0 \bigwedge \beta = 0, \,\gamma\in {\mathbb R}$\\
        \hline
        ${\mathcal{NG}}_{U\pm}$ & $\lp 0, \pm\frac{\sqrt{\gamma^2-2\beta\gamma-6}}{\gamma - 2\beta}, - \frac{\sqrt{6}}{2\beta - \gamma}, \sqrt{\frac{2\beta}{2\beta- \gamma}}, 0,0 \rp$ & $\beta, \lambda_1, \lambda_2 > 0 \bigwedge  \gamma \leq \beta - \sqrt{\beta^2 +6} $  \\
        \hline
        ${\mathcal{NG}}_{W\pm}$ or ${\mathcal K}_{\phi,\chi\pm}$ & $\lp \frac{\sqrt{6} (2 \beta + \gamma - \lambda_2)}{\lambda_1 (2 \beta - \lambda_2)}, \pm\frac{\sqrt{\lambda_2^2-2\beta\lambda_2-6}}{\lambda_2 - 2\beta}, - \frac{\sqrt{6}}{2\beta - \lambda_2}, 0,\sqrt{\frac{2\beta}{2\beta- \lambda_2}}, \pm 2 \sqrt{\frac{3 \beta (\lambda_2 - 2 \beta - \gamma)}{\lambda_1^2 (\lambda_2 - 2 \beta)^2}} \rp$ & $\lambda_1 > 0 \bigwedge \lambda_2 >\sqrt{6} \bigwedge \beta = 0 \bigwedge \gamma \neq 0 $  \\
        \hline
\end{tabular}
} \caption{Fixed points for the system in \eqref{eq:eomsDS} with exponential couplings and potentials.  Please refer to the main text for a detailed explanation of the existence conditions. }
    \label{tab:FPs_Exp}
\end{table}

\begin{table}[h]
   \centering
    \begin{tabular}{|c||c|c|c|c|c|}
        \hline
        \rowcolor{gray!30}
        \textbf{Point}  & $\Omega_\varphi$ & $\Omega$ & $w_\vp$ & $w_{\rm eff}$ & $\boldsymbol{\omega}$ \\ 
        \hline
                \hline
        \Kpm & $1$ & $0$ & $1$ & 1 & 0 \\
        \hline
       $\widetilde{\mathcal K}_- $& $1$ & $0$ & $1$  &$1$ & $0$ \\
        \hline
      $\widetilde{\mathcal K}_+$ & $1$ & $0$ &$1$ & $1$& $0$ \\
        \hline
        \F & $0$ & $1$  & $0$ & $0$ & $0$\\
        \hline
        $\widetilde{\mathcal F}_\pm$ & $0$ & $0$ &$0$ & $0$ & $0$\\ 
        \hline
        ${\mathcal S}_{\lambda_2}$  & $\frac{3}{\lambda_2^2}$ & $1-\frac{3}{\lambda_2^2}$ & $0$ & $0$& $0$\\
        \hline
        \Sg & $\frac{3}{\gamma^2}$ & $1-\frac{3}{\gamma^2}$ & $0$ &$0$ & $0$ \\
        \hline
        ${\mathcal G}_{\lambda_2}$  & $ 1$ & $ 0$ &  $ -1+\frac{\lambda_2^2}{3}$& $-1+\frac{\lambda_2^2}{3} $& $0$ \\
        \hline
        \Gg &  $ 1$& $0$ & $-1+\frac{\gamma^2}{3} $ & $-1+\frac{\gamma^2}{3} $& $0$ \\
        \hline
        $\mathcal{K_{\chi\pm}}$ & $1 $ & $ 0$  & $ 1$ & $1$ & $0$\\
        \hline
        ${\mathcal{NG}}_{U\pm}$ & $ 1 $ & $ 0$ & $1 +\frac{4\beta}{\gamma-2\beta}$ & $1+\frac{4\beta}{\gamma-2\beta}$  & $\pm\sqrt{6}\beta \frac{\sqrt{\gamma^2-2\beta\gamma-6}}{\gamma-2\beta}$\\
        \hline
        ${\mathcal K}_{\phi,\chi\pm}$  & $1 $ & $ 0 $ & 1 &1  & 0  \\
        \hline
\end{tabular}
\caption{Cosmological parameters for the fixed points in Table \ref{tab:FPs_Exp}. Note that $w_{\rm eff}$ is calculated for $w=0$. }
    \label{tab:CosmoParams_Exp}
\end{table}

\begin{enumerate}[{\bf 1.}]
    \item The points $\mathcal{K}_\pm$ and $\mathcal{F}$ correspond to the standard \emph{kination}- and \emph{fluid}-dominated fixed points, respectively, already present in the single-field quintessence scenario.  
For these three points we have $y_2 \propto e^{-\lambda_2 \phi} \rightarrow 0$, which, taking $\lambda_2 > 0$, implies $\phi \rightarrow +\infty$.  
This divergence of $\phi$ leads to two immediate requirements:  
(i) we must have $\beta \leq 0$ in order for the kinetic coupling $f$ to remain finite or vanish, and  
(ii) since we can always take $\lambda_1 \geq 0$, we require $\gamma \geq 0$ for $y_1$ to vanish.  
Note that $\lambda_1$ and $\gamma$ cannot vanish simultaneously, as otherwise $y_1$ would remain finite.  
Finally, because $y_f = \frac{y_1}{f} = 0$, we require $\gamma + 2\beta \geq 0$, which in turn cannot vanish at the same time as $\lambda_1$.
    \item The points $\widetilde{\mathcal{K}}_\pm$ and $\widetilde{\mathcal{F}}_\pm$ arise due to the introduction of the new variables $x_1$ and $y_f$, but physically correspond to the same regimes as above: \emph{kination} and \emph{fluid} domination.  
Their existence conditions are nearly identical to the previous ones, with the only difference being that $x_1$ and $y_f$ do not vanish at these points.  
In particular, the case $\beta = 0$ is now excluded, and consequently the condition $\gamma + 2\beta \geq 0$ is no longer required.

\item The points ${\mathcal S}_{\lambda_2}$ and ${\mathcal G}_{\lambda_2}$ correspond to the standard \emph{scaling} and \emph{scalar}-- (or \emph{geodesic}\footnote{We follow the convention and notation of \cite{Cicoli:2020cfj}, who refer to the scalar-dominated point as the \emph{geodesic} point.}) dominated solutions of the single-field case, with $w_{\rm eff}=0$ and $w_{\rm eff} = \frac{\lambda_2^2}{3}-1$, respectively.  
For these two points there are no additional constraints arising from the functional dependence of the dynamical variables on the fields.

\item The points $\mathcal{S}_{\gamma}$ and $\mathcal{G}_{\gamma}$ are analogous to the previous ones -- namely, \emph{scaling} and \emph{scalar}-dominated solutions -- but now associated to the potential coupling $g(\phi)$.  
Since the axion is non-dynamical ($x_1=0$), the potential coupling $g(\phi)$ effectively acts as a potential for the saxion.  
In contrast to the $\lambda_2$-associated points, here the condition $y_2=0$ requires $\phi \rightarrow +\infty$.  
Moreover, to ensure $y_f = 0$ we must have $f \rightarrow +\infty$, which in turn requires $\beta > 0$.  
Finally, for the point to be physical we still need $x_f = 0$, which implies that in this case $\dot{\chi}$ must vanish faster than $f$ grows in the $\phi \rightarrow +\infty$ limit.

\item  The points $\mathcal{G}_{\chi\pm}$ or $\mathcal{K}_{\chi\pm}$ appear at first sight to be a novel feature of the system, resembling  non-geodesic points  for the axion $\chi$ (since $x_f\ne0$, $\boldsymbol{w}\ne0$ at first sight).  
However, the condition $y_2 = 0$ implies $\phi \rightarrow +\infty$.  
Since here both $x_f$ and $y_f$ are non-zero, this requires $f$ to remain finite and non-vanishing.  
This is only possible if $\beta = 0$, i.e.~in the absence of kinetic coupling, and thus $\boldsymbol{w}=0$.
In that case, the points reduce to
\[
\left( \frac{\sqrt{6}}{\lambda_1},\, \pm1,\, 0,\, 0,\, 0,\, 0 \right),
\]
which correspond to a pure kination dominated points, associated with the axion\footnote{If one ignored the constraints arising from the functional form of the potentials and couplings—considering only the Friedmann constraint and the reality conditions—one would instead find the region 
$ 
\big( \lambda_1 > 0 \ \land\  \lambda_2 \geq 0 \ \land\  \beta < 0 \ \land\  \gamma \geq 0 \big) 
\ \ \lor\ \ 
\big( \lambda_1 > 0 \ \land\  \lambda_2 \geq 0 \ \land\  \beta > 0 \ \land\  \gamma \leq 0 \big),
$ 
which would misleadingly suggest the existence of a genuine geodesic point.}.

    \item  The points $\mathcal{N}\mathcal{G}_{U\pm}$ are  genuinely \textbf{\emph{non-geodesic}} (or multiscalar) solutions, with no analogue in the single-field case.  
Its non-geodesic parameter and effective equation of state are
\begin{equation}\label{eq:NGuws}
\boldsymbol{\omega} =\sqrt{6} \, \beta \,\frac{\sqrt{\gamma^2 - 2 \beta \gamma - 6}}{\gamma - 2 \beta} \,  \,, 
\qquad 
w_{\rm eff} = 1 + \frac{4\beta}{\gamma - 2\beta} \,.
\end{equation}
Thus, these points admit multifield accelerated expansion for the parameter range:\\
 $\gamma>-\beta \,,$  together with the existence condition $\gamma<\beta-\sqrt{\beta^2+6}$. 

From  $y_2 = 0$ we infer $\phi \to +\infty$.  
The conditions $y_1 \neq 0$, $y_f = 0$, $x_1 = 0$, and $x_f \neq 0$ together imply that $f \to +\infty$, which requires $\beta > 0$.  
When combined with the mathematical existence conditions above, this leads to $\chi \to +\infty$—formally a constant (since $x_1 = 0$) but with an infinitely large value\footnote{If one were to ignore the link between variables through their field dependence, focusing solely on the Friedmann constraint and the reality of the variables, the allowed region would instead be
\(
\big( \beta < 0 \ \land\  \gamma \geq \sqrt{\beta^2 + 6} + \beta \big) 
\ \ \lor\ \ 
\big( \beta = 0 \ \land\  (\gamma \leq -\sqrt{6} \ \lor\  \gamma \geq \sqrt{6}) \big),
\)
which would misleadingly suggest a broader set of possibilities.}. In subsection \ref{subs:NGusols} we explore this point in more detail.

    \item 
   The final points,  ${\mathcal{NG}}_{W\pm}$  or $\mathcal{K}_{\phi,\chi\pm}$ initially appear to be  promising non-geodesic solutions (again $x_{f,c}\ne0$ and thus $\boldsymbol{w}\ne0$).  
However, the condition $y_1 = 0$ and $y_f\ne0$ implies $f\to 0$ and therefore, $\phi \to \infty$ and thus $x_f\to 0$ and $y_2\to 0$,  which is not the case. Therefore we should set $y_f=0$, for which there are two possibilities: either $\beta=0$ or $(\lambda_2-2\beta-\gamma)=0$. In the former case, the points reduce to
\[
\left( -\frac{\sqrt{6}(\gamma - \lambda_2)}{\lambda_1 \lambda_2},  \pm\frac{\sqrt{\lambda_2^2 - 6}}{\lambda_2},  \frac{\sqrt{6}}{\lambda_2},  0,  0,  0 \right),
\]
which correspond to  \emph{multifield kination}-dominated solutions with $w_{\rm eff} = 1$, where both the axion and the saxion contribute to the total kinetic energy.
In the later case, $x_1$ vanishes and this implies that  $f$ has to diverge to have a non-vanishing $x_f$. This in turn implies that either $y_2= 0$ and we go back to the case $\beta=0$ above;  or $\lambda_2=0$, which is not possible as this makes $x_f\in {\mathbb{C}}$. 
\\
  
\end{enumerate}

\subsubsection{Stability of the fixed points}

In order to assess the cosmological relevance of a the fixed points of our system, we now  determine its stability. Stable attractors can represent generic late-time solutions, while unstable or saddle points describe transient behaviour. The standard approach to determine stability relies on linearising the system around each critical point and analysing the eigenvalues of the Jacobian. If all eigenvalues have non-vanishing real parts, the point is said to be \emph{hyperbolic}, and its stability is determined straightforwardly: negative (positive) real parts signal stability (instability). 

In our setup, however, we find at least two fixed points which feature one or more vanishing eigenvalues, while for other points, special values of the parameters make some of the eigenvalues vanish. 
Such points are \emph{non-hyperbolic}, and the linear analysis alone is inconclusive. To investigate their stability we  resort to the \emph{centre manifold theory} (CMT), which provides a systematic way to reduce the dynamics to the submanifold spanned by the zero eigenvalue directions, where the leading non-linear terms control the behaviour. Intuitively, the CMT allows us to focus only on the “slow” directions of the flow (those associated with zero eigenvalues), while the stable and unstable directions are integrated out. As explained for example in \cite{Bahamonde:2017ize,wiggins2003applied}, whenever at least one of the eigenvalue is zero and at least one of the non-zero eigenvalues is positive,  the point is unstable independently of the presence of additional vanishing eigenvalues, and no centre-manifold analysis is required.
Thus, while hyperbolic points can be classified immediately through linearisation, the fate of non-hyperbolic ones requires a more delicate treatment via centre manifold theory (or other methods) whenever the non-zero eigenvalues are all negative.

The eigenvalues of the fixed points for our system  are given  in Table \ref{tab:eigenv_Exp}, while their corresponding stability properties are summarised in Table \ref{tab:FPs_Exp_stab}. In the following, we discuss the stability of each fixed point in detail\footnote{For certain  parameter choices, the  stability analysis becomes  more involved. Since these cases do not correspond to specific physically relevant limits, and do not modify the qualitative behaviour of the system, we do not examine them further.
}. Note that for the points ${\mathcal S}_{\lambda_2}$, \Sg, \Gg, and ${\mathcal G}_{\lambda_2}$, we set $w = 0$.

 \begin{itemize}
      
    \item The kination points \Kpm \, can be either repellers or saddles for specific parameter regions as indicated in Table \ref{tab:FPs_Exp_stab}.  
    Moreover, for the values $\beta=0$, $ \lambda_2 = \sqrt{6}$ (for \Kp), $\gamma = \pm \sqrt6 $ and $\gamma = \pm \sqrt{6} - 2 \beta$ at least one of the eigenvalues vanishes. Using the  CMT as explained above  and,
     since  $3 (1- w)$ is always positive, the points in this region are unstable. 

    \item The fixed points $\widetilde{\mathcal{K}}_\pm$ are saddles for specific parameter regions (see Table \ref{tab:FPs_Exp_stab}). Moreover,     
    for   $\lambda_2 = \sqrt{6}$, the point 
    $\widetilde{\mathcal{K}}_+$ has one zero eigenvalue. Using again  the CMT, this point is found to be unstable. 

  \item The fluid point \F \, is a saddle whenever it exists. 
    
    \item The  points $\widetilde{\mathcal F}_\pm$ are  unstable because one of the eigenvalues is zero and the others have always mixed signs.

    \item The point  ${\mathcal S}_{\lambda_2}$  is  an attractor for a region of the parameter space, and a saddle for the regions indicated in Table \ref{tab:FPs_Exp_stab}. 
    In the following  special cases one eigenvalue vanishes while one of the non-vanishing eigenvalues is positive, so the points are unstable:  $\lambda_2=\sqrt{3}$ with $\gamma<\sqrt{3}$; $\lambda_2=-2\beta$ and $\lambda_2=-4\beta$ together with $|\beta|>\sqrt{3/14}$;  $|\beta|<\sqrt{3/14}$ with  $\gamma<6|\beta|$; and finally for $\lambda_2=2\beta+\gamma$.

    \item The scaling point  \Sg \, is an attractor for some parameter region as indicated in Table \ref{tab:FPs_Exp_stab}. On the other hand, for $\gamma=-2\beta, -4\beta, -\sqrt{3}$, one of the eigenvalues is zero with at least another non-vanishing eigenvalue being positive. Thus, in these cases, the point is unstable. Likewise, the point is unstable when $\lambda_2<\sqrt{3}$ in combination with either $\gamma=\sqrt{3}$ or $\gamma=\lambda_2$.\footnote{For the special case $\gamma=\lambda_2=\sqrt{3}$ two eigenvalues are zero and the CMT analysis gets more involved, but it can be checked numerically that in this case, the point is an attractor. }     
    \item The points ${\mathcal K}_{\chi\pm}$     are always unstable, because four of the eigenvalues are exactly zero but the other two are always positive. 
    \item The scalar–dominated solution \Gg \, can behave either as an attractor or as a saddle, depending on the parameters, as summarised in Table~\ref{tab:FPs_Exp_stab}. For certain isolated values, namely
\(
   \gamma = 2\beta - \sqrt{4 \beta^2 + 6}, 
   \quad \gamma = \beta - \sqrt{\beta^2 + 6}, 
\)
at least one eigenvalue vanishes with at least another non-vanishing eigenvalue positive, and the point is unstable according to the CMT. Similarly the point is unstable for   
$ \gamma = \sqrt{3}$ together with $\lambda_2<\sqrt{3}$, and 
   $\gamma = \lambda_2$ together with $\lambda_2>\sqrt{3}$.


\item The scalar–dominated point ${\mathcal G}_{\lambda_2}$  is likewise either an attractor or a saddle, depending on the region of parameter space, as summarised in Table~\ref{tab:FPs_Exp_stab}. For the special choices 
\(   \lambda_2 \in \{0, \gamma, 2\beta+\gamma, \sqrt{3}\},
\) 
at least one eigenvalue vanishes. In contrast to the previous case, where the CMT analysis revealed instability due to a positive eigenvalue, here the situation is more subtle: stability depends on higher–order terms along the centre manifold. 
To analyse this, one can transform the system into its normal form, so that the Jacobian becomes diagonal. This identifies the center manifold variables (those associated with zero eigenvalues). The remaining variables can then be expanded perturbatively in terms of the center ones, with coefficients determined order by order from the original equations. Substituting back yields an effective reduced system on the center manifold, which allows the stability to be determined (see section 2.5.3 of \cite{Bahamonde:2017ize} for an explicit example, and \cite{wiggins2003applied} for more details). Applying this method, we find: 
\begin{itemize}
  \item for $\lambda_2=0$, the point is stable;
  \item for $\lambda_2=\gamma < \sqrt{3}$, it is stable if $\beta=0$, otherwise unstable;
  \item for $\lambda_2=\gamma + 2 \beta < \sqrt{3}$ with $6 + 4 \beta^2 - \gamma > 0$, it is stable, otherwise unstable;
  \item for $\lambda_2=\sqrt{3}$, the point is unstable.
\end{itemize}
Instability arises also for $\lambda_2=\beta + \sqrt{6 + \beta^2}$.\footnote{For the specific value $\lambda_2=2\beta + \sqrt{6 + 4\beta^2}$ the analysis is again more involved.}
 
   \item The new points ${\mathcal{NG}}_{U\pm}$   exhibit an interesting  behaviour.  They lie on an invariant sub-manifold ($x_1 =  y_2 =y_f = 0$), so in the full phase space they  generically behave as a saddle.  Restricting the dynamics to the invariant sub–manifold, trajectories remain confined to it indefinitely, and in this reduced system the ${\mathcal{NG}}_{U\pm}$   points become  true (sub-)attractors whenever $\gamma > - 2 \beta$, while they are  saddles otherwise. This shows that our framework uncovers  genuine ${\mathcal{NG}}_{U\pm}$  attractors — albeit only within a sub–manifold. We illustrate this  behaviour in Subsection~\ref{subs:NGusols} and we return to the comparison with the flat–axion case in  Section~\ref{sec:flataxion}. Finally, note that for the special case $\gamma = - 2 \beta$ one eigenvalue vanishes, but the presence of positive eigenvalues renders the points unstable.

    \item The points $\mathcal{K}_{\phi,\chi\pm}$   have three  eigenvalues that vanish identically. For $\mathcal{K}_{\phi,\chi-}$, one eigenvalue is positive, rendering the point unstable.  In the case of $\mathcal{K}_{\phi,\chi+}$, although three eigenvalues are negative, the point corresponds to a multifield kination-dominated configuration and is therefore expected to represent an unstable point.

 \end{itemize}

\begin{table}[H]
    \centering
    \resizebox{17cm}{!}{
    \begin{tabular}{|c|c|}
        \hline
        \rowcolor{gray!30}
        \textbf{Point} & \textbf{Eigenvalues} \\
        \hline\hline
        \Kpm & $ \lp 3 (1 - w) , \mp 2\sqrt{6} \beta, \mp \sqrt{6} \beta, \frac{1}{2} \lp 6 \mp \sqrt{6} \gamma \rp , \frac{1}{2} \lp 6 \mp 2 \sqrt{6} \beta \mp \sqrt{6} \gamma \rp , \frac{1}{2} ( 6 \mp \sqrt{6} \lambda_2) \rp$ \\
        \hline
        $\widetilde{\mathcal{K}}_+$ & $ \lp 3 (1 - w), - \sqrt{6} \beta, \sqrt{6} \beta, - \sqrt{3} (\sqrt{2} \beta \pm \sqrt{ \beta (- \sqrt{6} + 4 \beta + \gamma)} \, ) , \frac{1}{2} (6 - \sqrt{6} \lambda_2) \rp$ \\
        \hline
        $\widetilde{\mathcal{K}}_-$ & $\lp 3 (1 - w), - \sqrt{6} \beta, \sqrt{6} \beta,  \sqrt{3} (\sqrt{2} \beta \pm \sqrt{ \beta ( \sqrt{6} + 4 \beta +\gamma)} \, ) , \frac{1}{2} (6 + \sqrt{6} \lambda_2) \rp$ \\
        \hline
        \F & $\frac{3}{2} \lp  1-w, w - 1,  w - 1, 1+w, 1+w, 1+w \rp$ \\
        \hline
        $\widetilde{\mathcal F}_\pm$ & $ \frac{3}{2} \lp 0, w - 1,  w - 1, 1+w, - \frac{1}{2} (1- w + \sqrt{(-1+w)(7+9 w)}), \frac{1}{2} \lp w -1 + \sqrt{(-1+w)(7+9 w)}\rp \rp$ \\
        \hline
        ${\mathcal S}_{\lambda_2}$ & $ \frac{3}{2} \lp 1-\frac{\gamma}{\lambda_2}, \, 1-\frac{2\beta + \gamma}{\lambda_2}, \, -\lp1 + \frac{2\beta}{\lambda_2 }\rp, \, -\lp1 + \frac{4\beta}{\lambda_2 }\rp, \,  - \frac{1}{2} \lp 1 \mp \frac{ \sqrt{24 - 7 \lambda_2^2}}{\lambda_2} \,\rp \rp$ \\
        \hline
        \Sg & $ \frac{3}{2} \lp - \frac{2 \beta}{\gamma}, \, - \lp 1 + \frac{2\beta}{\gamma}\rp , \, - \lp1 + \frac{4\beta} {\gamma} \rp, \, -\frac{1}{2} \left(1 \pm \frac{\sqrt{24 - 7 \gamma^2}}{\gamma}\right), \, 1 - \frac{\lambda_2}{\gamma}
        \rp$ \\
        \hline
        ${\mathcal G}_{\lambda_2}$ & $ \lp \frac{1}{2} \lambda_2(\lambda_2-\gamma), \, \frac{1}{2} \lambda_2 \lp\lambda_2-\gamma - 2 \beta \rp, \, \lambda_2^2 - 3, \, \frac{1}{2} (\lambda_2^2- 6), \, \frac{1}{2} (\lambda_2^2 -4\beta \lambda -6), \, \frac{1}{2}(\lambda_2^2 -2 \beta \lambda -6) \rp$ \\
        \hline
        \Gg & $\lp - \beta \gamma, \, \gamma^2-3, \, \frac{1}{2} (\gamma^2 -6), \, \frac{1}{2} (\gamma^2 -2 \beta \gamma -6), \, \frac{1}{2} (\gamma^2 -4 \beta \gamma -6), \frac12 \gamma(\gamma - \lambda_2) \rp$ \\
        \hline
        ${\mathcal K}_{\chi\pm}$  & $\lp 0,\, 0,\, 0,\, 0,\, 3,\, 3- 3 w \rp$\\
        \hline
        ${\mathcal{NG}}_{U\pm}$  & $ \lp \frac{6 \beta}{2 \beta - \gamma} , \, \frac{6 \beta}{2 \beta - \gamma}, - 3 \frac{2\beta + \gamma}{2\beta - \gamma}, \, \frac{-1}{2 \beta - \gamma} (3 \beta \pm \sqrt{3 \beta \, [8\gamma \beta^2 +(27-8\gamma^2)\beta +2\gamma(\gamma^2-6)]} \, ), \frac{3(\lambda_2 - \gamma)}{2\beta - \gamma} \rp$ \\
        \hline
        $\mathcal{K}_{\phi,\chi\pm}$ & $\lp 0, \, 0, \, 0, \, - \frac{3 (\gamma - \lambda_2)^2}{\lambda_1 \lambda_2}, - \frac{3 (\gamma - \lambda_2)^2}{\lambda_1 \lambda_2}, \mp \frac{6 (1-w)}{\lambda_2} \sqrt{6 - \frac{36}{\lambda_2^2}} \rp$ \\
        \hline
\end{tabular}
    }\caption{Eigenvalues of the fixed points in Table \ref{tab:FPs_Exp}. For ${\mathcal S}_{\lambda_2}$, \Sg, \Gg, ${\mathcal G}_{\lambda_2}$ we set $w = 0$.}
    \label{tab:eigenv_Exp}
\end{table}

\begin{table}[h]
  \hskip-0.7cm   \resizebox{18.5cm}{!}{
    \begin{tabular}{|c|c|c|c|}
        \hline
        \rowcolor{gray!30}
        \textbf{Point} & \textbf{Attractor} & \textbf{Repeller} & \textbf{Saddle} \\
        \hline\hline
        \Kp & Never & $\lambda_1 \geq 0 \bigwedge 0 \leq \lambda_2 < \sqrt{6} \bigwedge \beta<0 \bigwedge \gamma <\sqrt{6} $ & \multicolumn{1}{c|}{\begin{tabular}[c]{@{}c@{}c@{}} $\lambda_1, \lambda_2 \geq 0 \bigwedge \lambda_2, \gamma \neq \sqrt{6} \bigwedge \beta \neq 0  \bigwedge \gamma \neq \sqrt{6} - 2 \beta \bigwedge $ \\ $\lp \lambda_2 > \sqrt{6} \bigvee \gamma > \sqrt{6} \bigvee \beta>0 \bigvee \gamma > \sqrt{6} - 2 \beta \rp$ \end{tabular}} \\
        \hline
        \km & Never & $\lambda_1, \lambda_2 \geq 0  \bigwedge \beta>0 \bigwedge \gamma > - \sqrt{6} $ & \multicolumn{1}{c|}{\begin{tabular}[c]{@{}c@{}c@{}} $\lambda_1, \lambda_2 \geq 0 \bigwedge \gamma \neq -\sqrt{6} \bigwedge \beta \neq 0  \bigwedge \gamma \neq - \sqrt{6} - 2 \beta \bigwedge $ \\ $\lp \gamma < - \sqrt{6} \bigvee \beta<0 \bigvee \gamma < - \sqrt{6} - 2 \beta \rp$ \end{tabular}} \\
         \hline
         $\widetilde{\mathcal{K}}_+$ & Never & Never & $\lambda_1, \lambda_2 > 0  \bigwedge \gamma > \sqrt{6} - 2 \beta \bigwedge \beta < 0 \bigwedge \lambda_2 \neq \sqrt{6} $ \\
         \hline
         $\widetilde{\mathcal{K}}_-$ & Never & Never & $\lambda_1, \lambda_2 > 0  \bigwedge \gamma < - \sqrt{6} - 2 \beta \bigwedge \beta < 0 $ \\
         \hline
        \F & Never & Never & When it exists \\
        \hline
        $\widetilde{\mathcal F}_\pm$ & Never & Never & Never \\
        \hline
         ${\mathcal S}_{\lambda_2}$ & $\hbox{min}(\gamma, 2 \beta+\gamma)> \lambda_2>\sqrt{3} \bigwedge 4 \beta > -\lambda_2$ & Never & $\lp \lambda_2 > \hbox{min}(\gamma, 2 \beta+\gamma) \bigvee 4\beta < -\lambda_2 \rp \bigwedge \lambda_2 \neq \sqrt{3}, \gamma, 2 \beta + \gamma, -2\beta, -4 \beta$ \\
         \hline
         \Sg & $\lambda_1,\beta>0 \bigwedge \gamma > \sqrt{3} \bigwedge \lambda_2>\gamma$ & Never & $ \lp \gamma < - \sqrt{3} \bigwedge - \gamma \neq \beta, 2 \beta \rp \bigvee \lp \gamma>\sqrt{3} \bigwedge \lambda_2 < \gamma \rp $ \\
         \hline
         \Gg & $0 < \gamma < \hbox{min} (\sqrt{3}, \lambda_2) \bigwedge \lambda_2,\beta>0 \bigwedge \lambda_1 \geq 0 $ & Never & \multicolumn{1}{c|}{\begin{tabular}[c]{@{}c@{}c@{}} $\lb - \sqrt{6} < \gamma < 0 \bigwedge \gamma \neq \lp2b - \sqrt{4 \beta^2 - 6} \rp \bigwedge \gamma \neq \lp b - \sqrt{\beta^2 - 6} \rp \rb $ \\ $\bigvee \hbox{min} (\sqrt{3}, \lambda_2) < \gamma < \sqrt{6}$ \end{tabular}} \\
         \hline
         ${\mathcal G}_{\lambda_2}$ & $\lambda_1 \geq 0 \bigwedge 0 < \lambda_2 < \hbox{min }(\gamma ,\, \gamma + 2 \beta, \sqrt{3})$ & Never & $ \hbox{min }(\gamma ,\, \gamma + 2 \beta, \sqrt{3}) < \lambda_2 <\sqrt{6} $\\
         \hline
         ${\mathcal K}_{\chi\pm}$ & Never & Never & Never \\
         \hline
        ${\mathcal{NG}}_{U\pm}$ & Never & Never & $\beta,\lambda_1, \lambda_2 > 0 \bigwedge \gamma \leq \beta - \sqrt{\beta^2+6}$ \\
         \hline
         $\mathcal{K}_{\phi,\chi\pm}$ & Never & Never & Never \\
         \hline
\end{tabular}
    }\caption{Stability analysis summary for the fixed points in Table \ref{tab:FPs_Exp}.  }
    \label{tab:FPs_Exp_stab}
\end{table}


Most of the fixed points described above correspond to standard solutions with well-known properties in scalar cosmological models, namely kination, fluid domination, matter scaling, and geodesic (scalar-dominated) points. The genuinely new fixed points arising in the multifield case are the non-geodesic ones, which we therefore explore next. Given the six-dimensional nature of the phase space, a global numerical exploration of all standard fixed points would not provide additional insight beyond the analytic classification already presented, as these solutions are well understood in the literature. We thus focus in what follows on the genuinely new non-geodesic solutions.

\subsubsection{The   non-geodesic points,  ${\mathcal{NG}}_{U\pm}$  \,\,}\label{subs:NGusols}

As we have discussed above, the only non-geodesic points that arise in the case of exponential potentials are the ${\mathcal {NG}}_{U\pm}$  points, which turn out to be  attractors  on the subspace $(x_1,y_2,y_f)=(0,0,0)$ for $\gamma>-2\beta$. 
However, we emphasise that these non-geodesic fixed points 
are only  saddles in the full phase space (see Table \ref{tab:FPs_Exp_stab}). 
Therefore, if we fix the initial conditions   to the  subspace $(x_1,y_2,y_f)=(0,0,0)$, the system will stay on it for the full evolution and reach  ${\mathcal {NG}}_{U\pm}$   in the asymptotic future. We  show  this  explicitly in a concrete example below.

Note that when the non-geodesic points ${\mathcal{NG}}_{U\pm}$   exists, the scaling point \Sg\, also exists and it is a saddle. Moreover, also the fluid point $\cal F$, relevant for matter domination, exists and it is also a saddle (see Tables \ref{tab:FPs_Exp} and \ref{tab:FPs_Exp_stab}). 
Therefore, when ${\mathcal{NG}}_{U\pm}$   exist, there can  be two matter dominated epochs (recall we are not including radiation): one determined by the fluid point ${\mathcal F}$ and one by the matter scaling point \Sg.
Depending on the initial conditions, after kination, matter domination can either take place at the fluid point, moving to the scaling point to finally evolve  towards the non-geodesic (sub-)attractors: (${\cal K}_\pm \to {\cal F}\to {\cal S}_\gamma \to {\mathcal {NG}}_{U\pm}$ ); or pass far from the fluid point, taking place mostly at the  scaling point moving towards the non-geodesic (sub-)attractors (${\cal K}_\pm \to {\cal S}_\gamma \to {\mathcal {NG}}_{U\pm}$). 

\begin{figure}[h]
        \centering
       \includegraphics[width=0.48\linewidth]{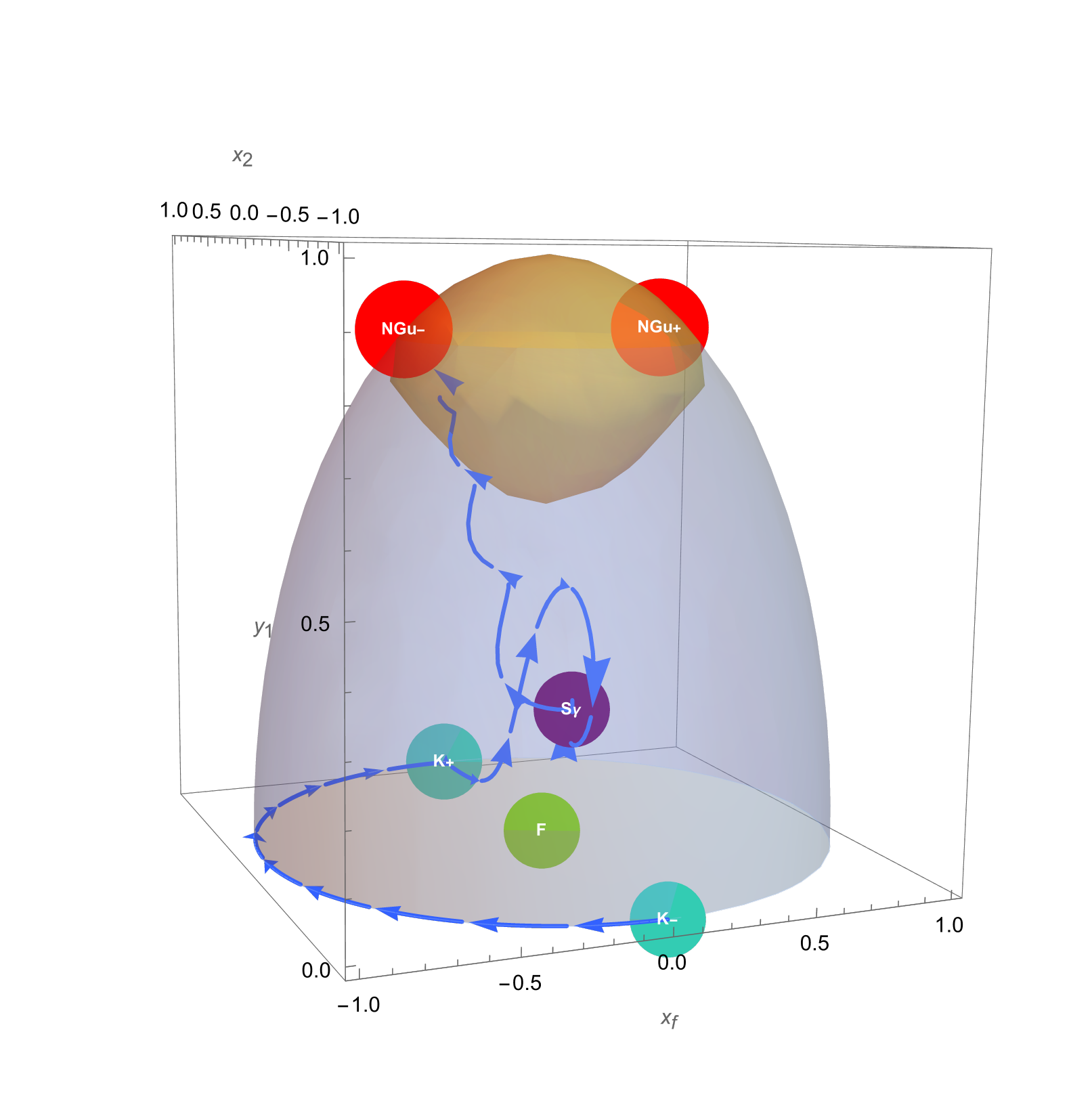}
          \includegraphics[width=0.48\linewidth]{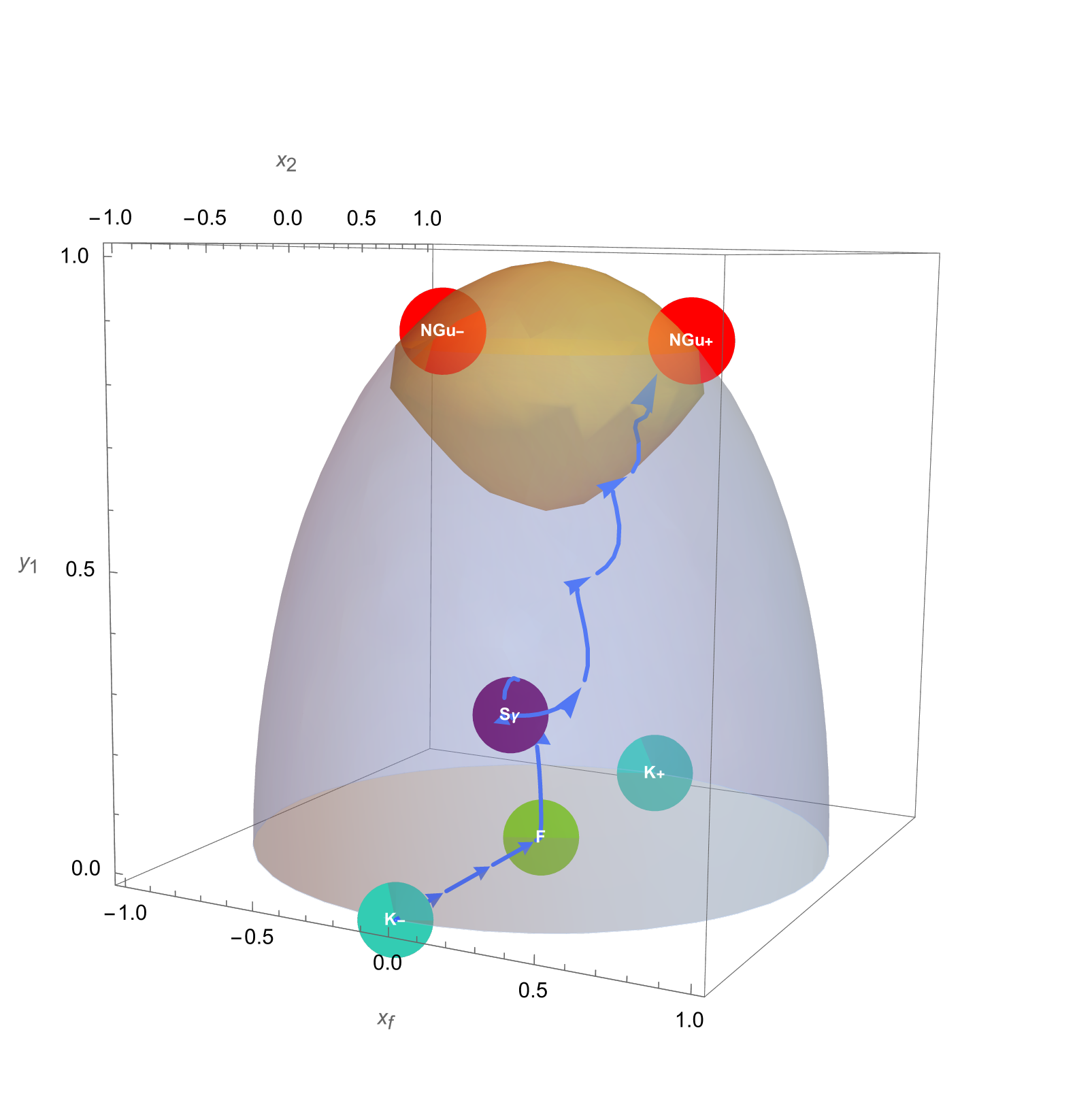}
        \caption{Phase space diagram for the 3D  subspace  $x_1=y_2=y_k=0$, for the parameters $(\beta,\gamma,\lambda_1,\lambda_2)=(8, -5, 3, 3)$, for which the  ${\mathcal {NG}}_{U\pm}$  points are (sub-)attractors as described in the main text. The yellow area denotes the region of the phase space where the universe is accelerating.
     \textbf{Left:} The diagram shows a trajectory with initial conditions $(x_1,x_f,x_2,y_1,y_2,y_f)=(0,-2\times10^{-10},-6\times10^{-8}, 2.6\times 10^{-6},0,0)$, starting in kination domination $(\mathcal{K_{\pm}})$, followed by the  scaling domination $(\mathcal{S_\gamma})$ and landing in the non-geodesic domination $\mathcal{NG}_{U-}$, in an accelerating universe.
    \textbf{Right:} The diagram depicts a trajectory with initial conditions $(x_1,x_f,x_2,y_1,y_2,y_f)=(0,2.5\times10^{-12},-0.2, 0.2,0,0)$, that starts at  kination domination $(\mathcal{K_{-}})$, followed by  fluid domination $(\mathcal{F})$, passing by a scaling epoch $\mathcal{S_\gamma}$ and ending at the non-geodesic  $\mathcal{NG}_{U+}$ point, in the accelerating region. }
    \label{fig:NGUPhasePl}
\end{figure}

We illustrate these two situations in Figure \ref{fig:NGUPhasePl}. We focus on examples where the ${\mathcal {NG}}_{U\pm}$  points allow for acceleration, which requires $\gamma>-\beta$ as discussed above. Interestingly, in this case, most of the parameter space gives rise to  spiral ${\mathcal {NG}}_{U\pm}$  (sub-)attractors, while the kination points are saddles, rather than repellers. 
Moreover, for the scaling point to be a possible candidate for (dark) matter domination, $\gamma$ needs to be sufficiently large. Assuming   $\Omega_{\rm dm}^{\rm max}\sim 0.84$ \cite{Rahimy:2025iyj}, gives $\gamma\gtrsim -4.33$. Choosing suitable values for  $\lambda_1, \lambda_2$, to avoid other fixed points to appear, we arrive at the following choice of parameters:  $(\beta,\gamma,\lambda_1,\lambda_2)=(8, -5, 3, 3)$. For this parameter choice the non-geodecisity parameter and effective equation of state, 
\eqref{eq:NGuws} are: 
\be
{\boldsymbol{\omega}}= \pm\frac{8\sqrt{66}}{7} \simeq\pm9.3\,, \qquad w_{\rm eff} = -11/21\simeq -0.52\,.
\ee

\subsection{Revisiting the shift symmetric axion }\label{sec:flataxion}

It is now useful to revisit the \emph{flat axion} case, where the shift symmetry of the axion remains fully unbroken, analysed in \cite{Cicoli:2020cfj} (see also \cite{Revello:2023hro,Seo:2024qzf,Shiu:2024sbe}), translating it into our set of variables.  %
This reformulation makes it possible to clearly identify the specific role played by the kinetic coupling\footnote{Recall that in the flat axion case there is neither a potential for the axion nor a potential coupling between the axion and the saxion. In our notation, this corresponds to $g = U = 0$.}.  
Within our framework, the flat axion case is obtained by setting $y_1 = 0$ (and therefore $y_f = 0$), and by dropping the equation for $y_1$ (and correspondingly $y_f$) from the dynamical system \eqref{eq:eomsDS}.
 That is, the system of equations reduces to:
\begin{subequations}\label{eq:eomsDSCicoli}
    \begin{align}
        x_1' &= 3 \, x_1 ( x_f^2 + x_2^2 - 1 ) - 2 \sqrt{6} \beta x_1 x_2 + \frac{3}{2} x_1 (1+w)\Omega \, ;\\
        x_f' & = 3 \, x_f ( x_f^2 + x_2^2 - 1 ) - \sqrt{6} \beta x_f \,x_2 + \frac{3}{2} x_f (1+w)\Omega \, ; \\
        x_2' & = 3 \, x_2 ( x_f^2 + x_2^2 - 1 ) + \sqrt{6} \beta x_f^2 + \sqrt{\frac{3}{2}} \lambda_2\, y_2^2 + \frac{3}{2} x_2 (1+w)\Omega \, ; \\
        y_2' & = \frac{\sqrt{6}}{2}  y_2 \, \lb - \lambda_2 x_2 + \sqrt{6} (x_f^2 + x_2^2) +\sqrt{\frac{3}{2}}  (1+w)\Omega \rb \, .
    \end{align}
\end{subequations}

\subsubsection{Shift symmetric axion fixed points and properties}

The fixed points of the system can be  readily obtained from the equations \eqref{eq:eomsDSCicoli} and they are summarised in Table~\ref{tab:fix_point_Cicoli}.  
From the table, we immediately see that, for all fixed points, $x_1=0$, as expected for a flat axion direction.
Physically, this reflects the fact that in the absence of a potential the axion $\chi$ remains frozen: with no external force acting on it, we have $\dot{\chi}=0$ identically.

\begin{table}[h]
    \centering
    \begin{tabular}{|c||c|c|c|c||c|}
    \hline
\rowcolor{gray!30}
        \textbf{Point}          & $x_1$ & $x_f$ & $x_2$ & $y_2$ &         \textbf{Existence conditions} \\
         \hline
         \hline
         $\mathcal{K}_\pm$& $0$ & $0$ & $\pm1$ & $0$ & $\lambda_2 > 0 \bigwedge \beta \leq 0 $ \\
         $\mathcal{F}$& $0$ & $0$ & $0$ & $0$ & $\lambda_2 > 0 \bigwedge \beta \leq 0 $\\
         $\mathcal{S}$& $0$ & $0$ & $\sqrt{\frac{3}{2}} \frac{1+w}{\lambda_2}$ & $\sqrt{\frac{3}{2}} \frac{\sqrt{1-w^2}}{\lambda_2}$ & $\lambda_2 \geq \sqrt{3(1 + w)} \bigwedge \beta \in \mathbb{R}$\\
         $\mathcal{G}$& $0$ & $0$ & $\frac{\lambda_2}{\sqrt{6}}$ & $\sqrt{1- \frac{\lambda_2^2}{6}}$ & $0 \leq\lambda_2 \leq \sqrt{6} \bigwedge \beta \in \mathbb{R}$\\
         \hline
         $\mathcal{NG}_\pm$& $0$ & $\pm\sqrt{\frac{\lambda_2^2- 2 \beta \lambda_2 - 6}{(2 \beta - \lambda_2)^2}}$ & $- \frac{\sqrt{6}}{2 \beta - \lambda_2}$ & $\sqrt{ \frac{2 \beta}{2 \beta - \lambda_2}}$ & $ \lambda_2 = \beta + \sqrt{\beta^2 +6} \bigwedge \beta \leq 0 $ \\
         \hline
    \end{tabular}
    \caption{Fixed points and existence conditions for the flat axion system \eqref{eq:eomsDSCicoli}. The points $\mathcal{NG}_\pm$ are included for completeness, although they are non-physical, as discussed in the main text. This is already reflected in their existence condition, since they reduce to the point $\mathcal{G}$ whenever the condition is satisfied.   }
    \label{tab:fix_point_Cicoli}
\end{table}

In this case, we recover the familiar fixed points discussed earlier: {\em kination} ${\mathcal K}_\pm$, {\em fluid} ${\mathcal F}$, {\em scaling} ${\mathcal S}$, {\em geodesic} (or scalar-dominated) ${\mathcal G}$, and the apparent {\em non-geodesic} points $\mathcal {NG}_\pm$, as obtained in \cite{Cicoli:2020cfj}.
However, once the kinetic coupling is disentangled by introducing the appropriate variables, it becomes clear that there is no region of parameter space where $\mathcal{NG}_\pm$ are simultaneously non-geodesic and physical. Indeed, this point is precisely the spurious point ${\bf (g)}$ listed in Appendix \ref{app:SFP}. As we discussed there, this point is not physical and does not correspond to a genuine physical  solution in the non-flat axion case either. 

Let us discuss the properties of the other  points in more detail.
We begin with the slightly unusual existence conditions for \Kpm\ and \F. The requirement $\beta \leq 0$ originates from the exponential dependence of $y_2 \propto e^{-\lambda_2 \phi}$, which vanishes only as $\phi \to +\infty$ (recall that by symmetry we take $\lambda_2 \geq 0$ and $\lambda_2 \neq 0$).
Because the kinetic coupling $f \propto e^{\beta \phi}$ shares the same exponential dependence, avoiding $f \to \infty$ requires $\beta \leq 0$.

In Table~\ref{tab:cosmoparam_flataxion} we list the values of the cosmological parameters $(\Omega_\varphi, \Omega, w_\varphi, w_{\rm eff})$ at the physical fixed points. As expected, these coincide with the results of \cite{Cicoli:2020cfj}.

\begin{table}[h]
    \centering
    \begin{tabular}{|c||c|c|c|c|}
    \hline
\rowcolor{gray!30}  
     \textbf{Point}   & $\Omega_\varphi$ & $\Omega$ & $w_\vp$ & $w_{\rm eff}$ \\
         \hline
         \hline
         \Kpm & $1$ & $0$ & $1$ & $1$ \\
         \F & $0$ & $1$ & undefined & $ w $ \\
         \S & $\frac{3}{\lambda_2^2}(1+ w)$ & $1- \frac{3}{\lambda_2^2}(1+ w)$ & $w$ & $w$ \\
         $\mathcal{G}$ & $1$ & $0$ & $\frac{\lambda_2^2}{3} - 1$ & $\frac{\lambda_2^2}{3} - 1$\\
         \hline
    \end{tabular}
    \caption{Cosmological parameters for the flat axion case's fixed points.}
    \label{tab:cosmoparam_flataxion}
\end{table}

\subsubsection{Stability of the shift symmetric axion fixed points}

We now turn to the linear stability analysis of the fixed points of the flat axion system \eqref{eq:eomsDSCicoli}.  
 The resulting eigenvalues and the associated stability properties are summarised in Tables~\ref{tab:flatAStab_eig} and \ref{tab:flatAStab_prop}.

\begin{table}[h]
   \centering
   \begin{tabular}{|c|c|}
   \hline
   \rowcolor{gray!30}
    \textbf{Point} &         \textbf{Eigenvalues} \\
   \hline
   $\mathcal{K}_+$ & $\lp -2 \sqrt{6}\,\beta,\, - \sqrt{6}\,\beta,\, \tfrac{1}{2}(6- \sqrt{6}\lambda_2),\, 3(1-w)\rp$ \\
   \hline
   $\mathcal{K}_-$ & $\lp 2 \sqrt{6}\,\beta,\, \sqrt{6}\,\beta,\, \tfrac{1}{2}(6+ \sqrt{6}\lambda_2),\, 3(1-w)\rp$ \\
   \hline
   $\mathcal{F}$ & $ \tfrac{3}{2} \lp w-1,\, w-1,\, w-1,\, w+1 \rp$ \\
   \hline
   $\mathcal{S}$ & $\lp - \tfrac{3}{2} - \tfrac{6 \beta}{\lambda_2},\, - \tfrac{3}{2} - \tfrac{3 \beta}{\lambda_2},\, 
    \tfrac{3}{4}\lp -1 \pm \sqrt{\tfrac{24-7\lambda_2^2}{\lambda_2^2}} \rp \rp$ \\
   \hline
   $\mathcal{G}$ & $\lp \tfrac12 (\lambda_2^2 - 4 \beta \lambda_2 - 6),\, \tfrac12(\lambda_2^2 - 2 \beta \lambda_2 - 6),\, 
   \lambda_2^2 - 3,\, \tfrac12(\lambda_2^2 - 6)\rp$ \\
   \hline
   \end{tabular}
   \caption{Eigenvalues of the flat axion system \eqref{eq:eomsDSCicoli}.}
   \label{tab:flatAStab_eig}
\end{table}

\begin{table}[h]
   \hskip-1cm\resizebox{18cm}{!}{
   \begin{tabular}{|c|c|c|c|}
   \hline
   \rowcolor{gray!30}
     \textbf{Point} & \textbf{Attractor} & \textbf{Repeller} & \textbf{Saddle} \\
   \hline
   $\mathcal{K}_+$ & Never & $\beta < 0 \wedge 0 < \lambda_2 < \sqrt{6}$ & $\lambda_2>\sqrt{6} \wedge \beta \leq 0$ \\
   \hline
   $\mathcal{K}_-$ & Never & Never & $\beta < 0 \wedge \lambda_2 > 0$ \\
   \hline
   $\mathcal{F}$ & Never & Never & $\beta \leq 0 \wedge \lambda_2 > 0$ \\
   \hline
   $\mathcal{S}$ & $\lambda_2 \geq -4\beta \wedge \lambda_2 > \sqrt{3}$ & Never & $-2\beta < \lambda_2 < -4\beta \wedge \lambda_2 > \sqrt{3}$ \\
   \hline
   $\mathcal{G}$ & $\lambda_2 \leq 2\beta + \sqrt{4\beta^2+6} \wedge 0 \leq \lambda_2 < \sqrt{3}$ & Never & 
   \begin{tabular}[c]{@{}c@{}} 
   $\big(\lambda_2 > 2\beta + \sqrt{4\beta^2+6} \wedge 0 \leq \lambda_2 < \sqrt{6}\big) \;\vee$ \\
   $\big(\sqrt{3} < \lambda_2 < \sqrt{6} \wedge \beta \in \mathbb{R}\big)$
   \end{tabular} \\
   \hline
   \end{tabular}
   }
   \caption{Stability properties of the flat axion fixed points.}
   \label{tab:flatAStab_prop}
\end{table}

As in the general case, certain parameter choices lead to one or more vanishing eigenvalues and thus to some non-hyperbolic critical points. We analyse these situations case by case:

\begin{itemize}
    \item {\em Kination points} \Kpm: These are never attractors, as expected.  
    For $\lambda_2 = \sqrt{6} \bigwedge \beta < 0 $, one eigenvalue of \Kp $\,$ vanishes, while for $\beta=0 \bigwedge\lambda_2 > 0$, both \Kpm $\,$ acquire two zero eigenvalues.  
    Applying the CMT, the presence of a positive eigenvalue $3(1-w)$ ensures that the points are unstable in these regions.  
    
     \item {\em Fluid point} \F: This point is generically a saddle, and never an attractor or a repeller.  
   
    \item {\em Scaling point} \S: This point is never a repeller.  
    For $ \lambda_2 =-2 \beta \bigwedge \lambda_2 \geq\sqrt{3} $, and for $ \lambda_2 =\sqrt{3} \bigwedge \beta \leq 0 \bigwedge \beta \neq - \frac{\sqrt{3}}{4}$, at least one eigenvalue vanishes.  
    In both cases, additional positive eigenvalues render the point unstable.

    \item {\em Scalar-dominated point} $\mathcal {G}$: This point is never a repeller.  
    For $ \lambda_2= \sqrt{3} \bigwedge\beta \in \mathbb{R}$, and for $\lambda_2 = 2 \beta + \sqrt{4 \beta^2 + 6} \bigwedge \beta > - \frac{\sqrt{3}}{4}$, at least one eigenvalue vanishes.  
    In these regions, the presence of positive eigenvalues implies that the point is unstable.  
\end{itemize}

In summary, rewriting the flat axion system of \cite{Cicoli:2020cfj} in terms of our variables makes it evident that the apparent non-geodesic point is not physical once the full field–variable relations are imposed.  
In contrast, in the more general framework considered in this work—where the axion has a potential and is kinetically coupled to the saxion—our analysis reveals  genuinely non-geodesic fixed points, $\mathcal{NG}_{U\pm}$.  
However, their stability is restricted to a lower-dimensional invariant submanifold of phase space, and they therefore cannot serve as  generic late-time attractors.

\subsection{Towards more general potentials and couplings}

Having determined the minimal set of variables required to close the dynamical system in the case of kinetic and potential couplings, it is natural to ask how the procedure extends to more general scalar potentials. Our framework makes this possible in a systematic way. In this subsection we illustrate the method for a simple but well-motivated example: a power-law potential for the axion, while retaining exponentials for the saxion.

A complete stability analysis of this case is considerably more challenging, as the phase space is seven-dimensional and all fixed points become non-hyperbolic. This would require a systematic centre manifold analysis for all fixed  points, which lies beyond the scope of the present work. We therefore defer the full stability analysis to future work, where we will also examine broader classes of axion and saxion potentials and kinetic coupling. Our aim here is instead twofold: (i) to demonstrate how our framework naturally extends beyond purely exponential potentials, and (ii) to highlight the type of dynamical structures that  arise in these more general settings.

As a concrete example, let us consider a more realistic axion potential, which   arises in string-inspired supergravity models, as we will discuss in the next section. Specifically, as often happens in multi-scalar string theory setups, we take exponential kinetic and potential couplings for the saxion, while adopting a quadratic potential for the axion, motivated by axion monodromy models\footnote{This can also be regarded as a Taylor expansion of the more common non-perturbative cosine potential.}  \cite{Silverstein:2008sg,Flauger:2014ana}. Concretely we consider:
\begin{equation} \label{pot_axion monodromy}
 f(\phi) = f_0\, e^{\beta \phi}\,,\qquad    
 V = W_0 \, e^{-\lambda_2 \phi} + \,  \frac{g_0\,e^{-\gamma \phi} M^2}{2}\chi^2  \,.
\end{equation}
For  $\beta=0$ and $\gamma = \lambda_2$, this case was studied in \cite{Marsh:2011gr,Marsh:2012nm} (for their $B=0$). 
Relative to the pure exponential case, $\lambda_1$ is no longer constant, and its evolution must be included via \eqref{eqn:eom_lambda1}, in addition to the equations for $(x_1,x_f,x_2,y_1,y_2,y_f)$. On the other hand, $\Gamma_1$ (see \eqref{eqn: Gammas def}) reduces to a constant\footnote{For a general power law $U=U_0 \chi^n$, one finds $\Gamma=(n-1)/n$.},
\(   \Gamma_1=  \frac{1}{2}\,.\)

\subsubsection{Fixed points in the power-law case}

Here we briefly outline the fixed point structure that arises in this case. A first striking feature is that all points have $x_1=0$, meaning that the axion is frozen. This is consistent with \cite{Marsh:2011gr,Marsh:2012nm}, even though in that case $\beta=0$ and $\gamma=\lambda_2$.  
Furthermore, the set of fixed points remains similar to the exponential case in Table~\ref{tab:FPs_Exp}, with some important differences. 

Specifically, we find the fixed points   \F, \Kpm, ${\mathcal S}_{\lambda_2}$ and  ${\mathcal G}_{\lambda_2}$. The axion can take arbitrary values at \F \, and \Kpm, while for ${\mathcal S}_{\lambda_2}$ and  ${\mathcal G}_{\lambda_2}$ it must be fixed at $\chi=0$ (so that $y_1=0$ and $y_2 \neq 0$). This implies $\lambda_1 \to \infty$, since $\lambda_1 \propto 1/\chi$. Apart from this restriction, the conditions on the remaining variables and parameters are the same as in the exponential case. 
There is one additional point that arises in the present case, namely
\[
 (x_1, x_f, x_2, y_1, y_2, y_f, \lambda_1) = 
\Big(0,0,\tfrac{\gamma}{\sqrt{6}},\sqrt{1-\tfrac{\gamma^2}{6}},0,0,\lambda_1\Big)\,,
\]
which requires $\gamma=0$. In that case the point reduces to $(0,0,0,1,0,0,\lambda_1)$ with $w_{\rm eff}=-1$, namely, a pure de Sitter point. 

The most remarkable differences with respect to the exponential case are i) the absence of the scaling and geodesic points \Sg, \Gg, and, more importantly, ii) of any consistent non-geodesic critical point. Indeed, all points with $x_f\neq0$ turn out to be unphysical. This indicates that this class of exponential-power-law potentials cannot support non-geodesic fixed points, even though full trajectories may still pass through non-geodesic regions. 

As remarked above, all the fixed points in this case are non-hyperbolic, since the eigenvalue associated to $\lambda_1$ vanishes identically. Standard linearisation is therefore insufficient, and centre manifold methods are required to assess stability. We leave this analysis, as well as the extension to more general potentials, to future work. In particular, when embedding these models in supergravity one typically encounters richer potential structures, where the situation may change. A systematic exploration of those cases will be pursued elsewhere.


\section{Realisations in string-inspired supergravity}\label{sec:sugra}

In Section \ref{sec:DS}, we analysed the axion--saxion system as a laboratory for multifield dynamics, with a particular emphasis on fixed points featuring a non-vanishing non-geodesic parameter $\boldsymbol{w}_c$. To this end, we first studied the case of constant parameters ($\beta, \gamma, \lambda_1, \lambda_2$), corresponding to exponentials, and then extended the discussion to the more realistic case of a quadratic (power-law) axion potential.  

We now turn to the question of how such potentials can arise from string-inspired supergravity constructions. This step is important for two reasons: first, it allows us to connect the dynamical systems approach to concrete models motivated by high-energy theory; and second, it illustrates how to systematically embed and test candidate potentials within our framework. As a case study, we focus on axion-monodromy--like potentials, in which the axion acquires a quadratic potential when its shift symmetry is broken at tree level by fluxes or other mechanisms.  

The scalar potential in $\mathcal{N}=1$ supergravity is constructed from the K\"ahler potential, $\k(\Phi_i, \bar\Phi_{\bar{\imath}})$, a real function of the chiral superfields $\Phi_i$, and the holomorphic superpotential $\w(\Phi_i)$, via the standard expression:
\begin{equation}
V = e^{\k/M_{\rm Pl}^2} \left( \k^{i\bar{\jmath}} D_i \w \, \overline{D_j \w} - 3 \frac{|\w|^2}{M_{\rm Pl}^2} \right),
\end{equation}
where $D_i \w \equiv \partial_i \w + \frac{1}{M_{\rm Pl}^2} (\partial_i \k) \w$, and $\k_{i\bar{\jmath}} \equiv \partial_i \partial_{\bar{\jmath}} \k$ defines the K\"ahler metric. 

\subsection{Constructing the supergravity potential}

We follow the approach of \cite{Kawasaki:2000yn,Kallosh:2010xz,Ferrara:2014kva}, and work with two ``orthogonal'' chiral superfields \cite{Roest:2013aoa}: a nilpotent goldstino superfield $S$, which satisfies a nilpotency condition $S^2 = 0$, and a superfield $\Phi$ containing the axion--saxion complex scalar (we denote both the superfields and their scalar components by the same letters for simplicity). 

Nilpotent superfields have been discussed recently in the context of string theory flux compactifications, where it has been shown \cite{Ferrara:2014kva} that using a nilpotent superfield\footnote{In a nilpotent chiral multiplet the scalar component, the sgoldstino, is a bilinear combination of the fermions. The auxiliary field of the nilpotent multiplet does not vanishing, and thus supersymmetry is broken spontaneously.}, one can reproduce the anti-D3-brane uplift term in the KKLT scenario \cite{Kachru:2003aw}\footnote{In \cite{Kallosh:2015nia} it was shown how the spectrum of a single anti-D3-brane in four-dimensional orientifolded IIB string models reproduces precisely the field content of a nilpotent chiral superfield with the only physical component corresponding to the fermionic goldstino. Moreover, the non-linear supergravity theory corresponding to this anti-D3 uplifting, including dependence on the bulk moduli, was computed in \cite{GarciadelMoral:2017vnz}.}. 
Moreover, coupling a nilpotent chiral superfield to supergravity leads to what is known as pure dS supergravity \cite{Bergshoeff:2015tra,Hasegawa:2015bza}, a theory without scalar degrees of freedom that naturally realises de Sitter solutions with non-linear supersymmetry (see also \cite{Bandos:2015xnf,Bandos:2016xyu,Nagy:2019ywi,Bansal:2020krz,Bansal:2024pgg} for alternative approaches).

Guided by these considerations, we take the superpotential of the form 
\begin{equation}
\w = S F(\Phi),
\label{eq:superpotentialAssumption}
\end{equation}
together with a general K\"ahler potential $\k(\Phi, \bar\Phi; S, \bar S)$, invariant under $S \to -S$. This symmetry ensures that $\k$ depends only on $S\bar S$ (and possibly $S^2 + \bar S^2$, though these vanish identically due to nilpotency). 

With these assumptions, one finds
\begin{equation}
D_S \w = F(\Phi), \qquad D_\Phi \w = 0,
\end{equation}
evaluated at $S=0$, and the scalar potential reduces to
\begin{equation} \label{eqn: Potential_from_S}
V = e^{\k(\Phi,\bar\Phi)/M_{\rm Pl}^2} \, \k_{S\bar{S}}^{-1}(\Phi,\bar\Phi) \, |F(\Phi)|^2.
\end{equation}
Our strategy is  therefore  to choose $\k$ and $F(\Phi)$ such that the resulting supergravity potential reproduces the structure of the effective scalar Lagrangian in \eqref{eq:scalarL}. In this way, we can embed the phenomenologically motivated potentials discussed earlier into a consistent supergravity framework, while keeping an eye on possible generalisations for future work.

\subsection{Functional Requirements for \texorpdfstring{$\k$}{K} and \texorpdfstring{$F(\Phi)$}{F(Phi)}}

Matching the structure of the scalar lagrangian \eqref{eq:scalarL} imposes the following functional forms:
\begin{enumerate}[(i)]
    \item \textbf{Kähler potential\footnote{The dependence on $\Phi$ resembles that found in \cite{GarciadelMoral:2017vnz} for the K\"ahler moduli found for non-linear supergravity theory corresponding to the anti-D3 uplifting. See also \cite{McDonough:2016der} for related work in the context of supergravity.}:}
    \begin{equation}
    \k = h(\Phi + \bar{\Phi}) + \mathcal{H}(\Phi + \bar{\Phi}) \, \frac{S \bar{S}}{M_{\rm Pl}^2},
    \end{equation}
    where $[h] = [\mathcal{H}] = 2$ in mass dimension, and $\Phi = \varphi + i \chi$ contains the saxion $\varphi$, and axion $\chi$. 
    
    \item \textbf{Superpotential:}
    \begin{equation}
    |F(\Phi)|^2 = W(\varphi) +g(\varphi) \, U(\chi),
    \end{equation}
    such that $[F] = 2$. This functional form allows us to match the separable structure used in our field-theoretic analysis.
\end{enumerate}

In addition to these constraints, the Kähler potential must also lead to a consistent kinetic structure upon going to component fields. This places further conditions on the form of $h(\Phi + \bar\Phi)$, which we discuss below.

With this choices for $\k, \w$, the supergravity kinetic term is given by 
\be
\mathcal{L}_{kin} = - \k_{\Phi \bar{\Phi}} \, \partial_\mu \Phi \, \partial^\mu\bar{\Phi} =-\frac14\frac{\partial^2 h}{\partial\vp^2} \lb \lp \partial \vp \rp^2 + \lp \partial \chi \rp^2 \rb. 
\ee
In order to obtain a kinetic term of the form $\mathcal{L}_{kin} \propto \lb \lp \partial \phi \rp^2 + f^2(\phi) \lp \partial \chi \rp^2 \rb$ we need 
  \begin{itemize}
                \item $\dfrac{\partial^2 h}{\partial\vp^2}>0$;
                \item $\dfrac{\partial^2 h}{\partial\vp^2} = \lp \dfrac{\partial q}{\partial\vp} \rp^2$;
                \item $q$ is invertible.
            \end{itemize}
    
            With these hypothesis we can define $d\phi = \dfrac{\partial q}{\partial\vp} d\vp$, such that  $\phi = q(\vp) $ and we can define $f(\phi) = \dfrac{\partial q}{\partial\vp} \lp \vp\lp q^{-1}(\phi) \rp \rp$ to obtain the desired form of the kinetic term.
Meanwhile, the scalar potential for $\Phi$ becomes 
\be
V = \frac{e^{h(\Phi+\bar\Phi)/M_{\rm Pl}^2}}{\mathcal H(\Phi+\bar\Phi)}\, |F(\Phi)|^2 M_{\rm Pl}^2  .
\ee

\subsubsection{A concrete example}

So far we have outlined the most  general form of the superpotential and the K\"ahler potential and the  constraints on the relevant functions such that we obtain a kinetic and potential coupling of the form described in the previous sections. Although in general there are several forms one can choose for $h$ and $F$ within string inspired models (while no particular constrain on $\mathcal{H}$ appeared) we wanted to find suitable functional forms such that the $\Gamma_i$ parameters defined in \eqref{eqn: Gammas def} remain constant. 
The choice is the following\footnote{In \cite{Aragam:2021scu}, a related model was studied for $n=1$, $F=a\Phi+b$, and $\alpha \to 3\alpha$, which admits sustained non-geodesic trajectories for small $\alpha$. In this case, however, the saxion potential is not a single exponential but instead takes the form 
$W(\phi)=W_0\,e^{-\lambda \phi}(b+a\,e^{\phi/\lambda})^2$, and therefore requires a separate analysis.} (we have set back $M_{\rm Pl}=1$):
\be
F=a \,\Phi \,,\qquad  h = -\alpha \ln \lp \Phi+\bar\Phi \rp\,, \qquad  \mathcal{H} = \lp \Phi+\bar\Phi \rp^n 
\ee
With these choices, we arrive at
\begin{subequations}
\begin{align}
    f(\phi) & = \sqrt{\frac{\alpha}{2}}\,e^{-\sqrt{2/\alpha}\,\phi}\,,\\
    W(\phi) &= 2^{-n-\alpha} a^2 e^{(2-n-\alpha)\sqrt{2/\alpha}\,\phi} \,,\\
    g(\phi) &=  e^{-(n+\alpha)\sqrt{2/\alpha}\,\phi} \,, \\
    U(\chi) &= 2^{-n-\alpha}  a^2 \chi^2\,.
    \end{align}
\end{subequations}
This implies that $\beta, \gamma,\lambda_2$ and $\Gamma_1$ are constants, given by:
\be
\beta=-\sqrt{2/\alpha}\,,\qquad \gamma=(n+\alpha)\sqrt{2/\alpha}\,,\qquad \lambda_2= (n+\alpha-2)\sqrt{2/\alpha} \,,\qquad \Gamma_1=\frac{1}{2}\,.
\ee
We thus find that this construction reproduces the class of power-law axion potentials analysed in the previous section, thereby embedding our dynamical-systems analysis into a supergravity setting. 
This approach  opens the door to more general possibilities, where richer multifield dynamics and non-geodesic effects may emerge -- an avenue we leave for future work.


\section{Discussion and Outlook}\label{sec:conclusion}

In this work, we have studied a system of two scalar fields interacting through both kinetic and potential couplings. Our focus was on an axion--saxion system, which arises naturally in string compactifications: saxions correspond to geometric moduli, while axions descend from higher-dimensional $p$-forms\footnote{See \cite{Cicoli:2023opf} for a recent review on string cosmology.}. In such setups the two fields typically combine into a complex scalar belonging to a chiral multiplet in $\mathcal{N}=1$ supergravity. 

Our main achievement has been to extend the dynamical systems toolkit for such models. By introducing the additional variables $x_1$ and $y_f$, we were able, not only to close the system and analyse in detail the stability of its fixed points, but also to disentangle the respective roles of the kinetic coupling and the axion velocity. 
 In this way, we incorporated not only the kinetic coupling---associated to a non-trivial field-space metric and widely studied in the literature---but also a potential coupling, which is a generic but so far largely unexplored  in this class of models. To the best of our knowledge, this represents the \emph{first systematic dynamical-systems treatment} of scalar theories with simultaneous kinetic and potential couplings. This generalisation provides a unified language for analysing multifield interactions and their cosmological implications.

Within this framework, we uncovered a pair of genuine \emph{non-geodesic} fixed points in the case of exponential couplings, denoted $\mathcal{NG}_{U\pm}$, thereby establishing the existence of consistent non-geodesic (sub-)attractors. As we showed, however, these points are attractors only within a restricted submanifold of the full system. At the same time, we revisited the flat axion case and demonstrated that the apparent non-geodesic fixed point found previously is in fact inconsistent once the full dynamics is taken into account. 

A further key result of our analysis was the derivation of a compact and general expression for the non-geodesicity parameter (turning rate) at the fixed points, valid for arbitrary couplings \eqref{eq:wc}:
\be
 \boldsymbol{w}_c = \sqrt{6}\,\beta\, x_{f,c} \, ,
\ee
where recall that $x_f = f \dot\chi/(\sqrt{6}H)$ and $\chi$ denotes the axion. This simple expression highlights three striking features: 
\begin{itemize}
    \item the non-geodesicity parameter depends directly on the non-trivial metric coupling $\beta$,
    \item it can be non-zero irrespective of the field-space curvature, and 
    \item  it is independent of the saxion kinetic energy $x_2$ and  potential energies $y_i$, though 
 the potentials remain important for the consistency conditions\footnote{See e.g.~the discussion about points $\mathcal{NG}_{W\pm}$.}.

\end{itemize}
This result is of independent interest, as it can be  applied not only to scalar dark energy but also to multifield inflationary scenarios, where non-geodesic trajectories play a central role. In fact, the two classes of non-geodesic trajectories identified in the literature---characterised by $x_2=0,\text{\,const.}$~\cite{Brown:2017osf,Garcia-Saenz:2018ifx,Aragam:2021scu}---fit naturally within our framework.

We also showed how our construction extends naturally to more realistic string-inspired potentials, such as power-law axion potentials combined with exponential saxion couplings. As a proof of principle, we embedded this structure into an explicit supergravity model, thereby demonstrating that our field-theoretic analysis can be consistently realised in a UV-motivated setting. This provides a concrete starting point for exploring richer classes of potentials and the associated cosmological dynamics.

In summary, we have presented a unified dynamical-systems framework for multifield models with both kinetic and potential couplings. Our results reveal new classes of fixed points, establish clear diagnostics of non-geodesicity, and pave the way towards a more systematic understanding of multifield dark energy and inflation. Looking ahead, it will be important to extend our framework to more general potentials motivated by string theory,  to test the observational consequences of non-geodesic dark energy, and to assess the perturbative stability of cosmologically viable background solutions. We expect that the tools developed here will serve as a foundation for future explorations of multifield cosmological dynamics, both in the late universe and during inflation.

\newpage
\section*{Acknowledgements}
We thank Susha Parameswaran for useful   discussions. DL thanks dipartimento di fisica Augusto Righi, Universit\`a degli studi di Bologna and 
INFN Sezione di Padova, in particular GSS project, for financial support.
The work of SR and IZ is partially supported by STFC
grants ST/T000813/1 and ST/X000648/1. 
\\
For the purpose of open access, the authors have applied a Creative Commons Attribution
license to any Author Accepted Manuscript version arising. Data access statement: no
new data were generated for this work.


\appendix 

\section{Spurious Fixed Points}\label{app:SFP}

In this appendix, we collect fixed points that appear when solving the system \eqref{eq:eomsDS} for fixed points, but turn out to be spurious. After physical and mathematical consistency conditions are imposed, these solutions can be seen to either reduce to those already presented in Table \ref{tab:FPs_Exp}, or else turn out to be unphysical. For completeness, we present them here together with a detailed explanation.

\begin{enumerate}[{\bf a)}]
    \item The point 
\[
    \lp 0,0,0, \sqrt{\tfrac{\lambda_2}{\lambda_2 - \gamma}}, \sqrt{\tfrac{\gamma}{\gamma - \lambda_2}}, 0 \rp\,,
\]
is consistent only in the special cases $\gamma = 0$ or $\lambda_2 = 0$. In these limits, it reduces respectively to the points $\mathcal{G}_\gamma$ and ${\mathcal G}_{\lambda_2}$. In either case the total equation of state becomes $w_{\rm eff}=-1$ \,,
corresponding to a cosmological constant. Since we are interested in non-zero couplings, we do not consider this point further.

    \item The points 
    {\footnotesize $$ 
       \sqrt{\tfrac{3}{2}} \lp  \frac{(1+w)(\lambda_2 - 2 \beta - \gamma)}{\lambda_1 \, \lambda_2}, 0, \frac{(1+w)}{ \lambda_2}, 0,   \frac{\sqrt{1-w^2}}{ \lambda_2}, \pm  \frac{\sqrt{(1+w)(\lambda_2 - 2 \beta - \gamma)\,[ 4(1+w) + \lambda_2(1- w) ]}}{\lambda_1 \, \lambda_2}\rp \, ,
    $$}
with $w_{\rm eff}=w$, can be absorbed into the scaling point $\mathcal{S}_\beta$. Indeed, the only way these points are consistent is in the region of parameter space where $x_1=0$, in which case they reduce exactly to $\mathcal{S}_\beta$.

    \item A similar argument to the point above applies to the points 
    \[\lp \frac{\lambda_2 (\lambda_2 - 2 \beta - \gamma)}{\sqrt{6} \, \lambda_1}, 0, \frac{\lambda_2}{\sqrt{6}}, 0, \sqrt{1- \frac{\lambda_2^2}{6}}, \pm \frac{\sqrt{\lambda_2 (\lambda_2 - 2 \beta - \gamma)(6+4 \beta \lambda_2 - \lambda_2^2)}}{\sqrt{6} \, \lambda_1} \rp\,,\] 
    which  then  reduce to ${\mathcal G}_{\lambda_2}$ with  $w_{\rm eff}=\frac{\lambda_2^2}{3}-1$.
     \item The points 
    $$ 
        \lp \frac{\sqrt{6}}{\lambda_1}, \pm\sqrt{\frac{\gamma(1-w)}{2(1+w)\beta + (1-w)\gamma}}, 0, \sqrt{\frac{2 \beta(w-1)}{2(1+w)\beta + (1-w)\gamma}}, 0, 0 \rp \, ,
    $$
     is physically consistent for $\beta=0$, in which case it reduces to  $\mathcal{K_\chi}$. 
    \item The point 
    $$
        \lp \frac{\sqrt{6} (\gamma -\lambda_2)}{\lambda_1(4\beta-\lambda_2)}, 0, - \frac{\sqrt{6}}{4 \beta-\lambda_2}, \sqrt{\frac{[4(1+ w) \beta + \lambda_2 (1-w)] \,  (6+4 \beta\lambda_2 - \lambda_2^2)}{(1+w) \, (4\beta-\lambda_2)^2 \, (\lambda_2-\gamma)}} \rp\,,
    $$ 
    is not physical. For it to be physically relevant, one would need to impose either $x_1=0$ or $y_1=0$. The first condition is never satisfied, while the second can hold in certain regions of parameter space. However, even then, consistency further requires either $\lambda_2=0$ or $y_2=0$. These two conditions can only be met if $\beta=0$, but this is incompatible with the requirement $f=0$. 

    \item The point 
    $$
       \lp \frac{\sqrt{6}}{\lambda_1}, 0, \sqrt{\frac{\gamma(1-w)}{2(1+w)\beta + (1-w)\gamma}}, 0, \sqrt{\frac{2\beta \, (w-1)}{2(1+w)\beta + (1-w)\gamma}} \rp\,,
    $$
    makes sense only for $\beta=0$, in which  case it is exactly $\mathcal{K_\chi}$.

\item The points 
$$ \lp0,\pm\sqrt{\frac{\lambda_2^2- 2 \beta \lambda_2 - 6}{(2 \beta - \lambda_2)^2}}, - \frac{\sqrt{6}}{2 \beta - \lambda_2}, 0,\sqrt{ \frac{2 \beta}{2 \beta - \lambda_2}},0\rp\,.
$$
These points arise in the flat axion case, as $y_1=0$ (as well as $y_f=0$), and look like  promising non-geodesic points with $x_f\ne0$.
It has $x_1=0$ but $x_f \neq 0$, implying $\dot{\chi}=0$ together with $f \to \infty$.
The latter condition forces $\phi \to \pm \infty$, which in turn drives $y_2 \to 0$ or $y_2 \to \infty$, unless $\lambda_2=0$.
But since $y_2 \neq 0$ for this point, and its existence condition forbids $\lambda_2=0$, this option is excluded.
Alternatively, forcing $x_f=0$ is possible if $\lambda_2 = \beta + \sqrt{\beta^2+6}$; however, in that case the point reduces exactly to the flat axion geodesic point ${\mathcal G}_{\lambda_2}$.
Finally, setting $y_2=0$ by choosing $\beta=0$ is also incompatible with $x_1=0$ and $x_f \neq 0$.\footnote{Technically, $\beta=0$ can still lead to a consistent solution if we also impose $x_f=0 \Rightarrow \lambda_2=\sqrt{6}$, in which case the point coincides with \Kp.}
We therefore conclude that this point  does not correspond to a genuine physical solution.

\item The points $(A_1,0,A_3,A_4,0,0)$ where 

    {\footnotesize
    \begin{subequations}
        \begin{align*}
            A_1 = & \frac{\sqrt{6}}{\lambda_1} + \frac{(4 \beta - \gamma)(4 \beta \gamma \pm \sqrt{2 \gamma} \sqrt{12 (1- w^2) \beta + 3 (1-w)^2 \gamma + 8 \beta^2 \gamma} \, )}{ \sqrt{6} \left[ 4 (1+w) \beta + \gamma(1-w) \right] \, \lambda_1} \, , \\
            A_3 = & \frac{ 2 \sqrt{6} \beta \gamma \pm \sqrt{3 \gamma} \sqrt{12 (1- w^2) \beta + 3 (1-w)^2 \gamma + 8 \beta^2 \gamma} }{ 3 \left(4 (1+w) \beta + \gamma(1-w)\right) } \, , \\
           A_4= & \sqrt{ \frac{ \pm \beta \left[ 3 (1-w)^2 \gamma + 16 \beta^2 \gamma + 4 \beta \left( 3(1-w^2) \pm \sqrt{2 \gamma} \sqrt{12 (1- w^2) \beta + 3 (1-w)^2 \gamma + 8 \beta^2 \gamma} \right) \right] }{ 3 \lb 4 (1+w) \beta + \gamma(1-w) \rb^2 } } \, , \\
        \end{align*}
    \end{subequations}
    }
    are not physically consistent as $x_f=0$ implies $f=0$, while $y_1\ne0$, but $y_f=0$, which is inconsistent. 

    \end{enumerate}

\section{Kinetic coupling as a dynamical variable}\label{app1}

In this appendix we discuss an alternative set of dynamical variables which, despite their apparent naturalness, do not turn out to be suitable for uncovering the features of the full axion--saxion system we consider, eq.~\eqref{eq:eomscalars}. As emphasised in the main text, when both kinetic and potential couplings are present, an additional variable is required in order to close the system. An apparently natural choice is to promote the kinetic coupling $f(\phi)$ itself to a dynamical variable. 

That is,  consider the set
\begin{equation}
   z=f(\phi)\,, \quad x_1=\frac{\dot\chi}{\sqrt{6} H}\,, \quad x_2=\frac{\dot\phi}{\sqrt{6} H}\,, \quad y_1=\frac{\sqrt{g(\phi)U(\chi)}}{\sqrt{3}H}\,, \quad y_2=\frac{\sqrt{W(\phi)}}{\sqrt{3}H}\,, \quad \Omega = \frac{\rho}{3 H^2}\,,
\end{equation}
together with
\begin{equation}
   \lambda_1=-\frac{U_\chi}{U}\,, \quad \lambda_2=-\frac{W_\phi}{W}\,, \quad \gamma= -\frac{g_\phi}{g}\,, \quad \beta =\frac{f_\phi}{f}\,.
\end{equation}
The Friedmann constraint \eqref{eq:Friedmann} then reads
\begin{equation}\label{eq:FriedConstr2}
   1 = z^2 x_1^2  + x_2^2 + y_1^2 + y_2^2 + \Omega\,.
\end{equation}

Using these variables, the equations of motion \eqref{eq:fulleoms} can be recast into a system of first-order differential equations for $z', x_1', x_2', y_1', y_2'$, namely
\begin{subequations}
\label{eqn: new dyn sys eqn}
\begin{align}
   x'_1 &= -3x_1 - 2\sqrt{6}\,\beta x_1 x_2 + 3x_1 x_2^2 + 3 z^2 x_1^3 + \frac{3}{2} (1+w_i)\Omega \,x_1 + \frac{3}{\sqrt{6}} \lambda_1 \frac{y_1^2}{z^2}\, , \\
   x'_2 &= -3x_2 + \sqrt{6}\,\beta z^2 x_1^2 + 3x_2^3 + 3 z^2 x_1^2 x_2 + \frac{3}{2} (1+w_i)\Omega \,x_2 + \frac{3}{\sqrt{6}}\left(\lambda_2 y_2^2 + \gamma y_1^2\right)\, , \\
   y'_1 &= \frac{y_1}{2} \left[ 6\left(x_2^2 + z^2 x_1^2\right) - \sqrt{6}\left(\gamma x_2 + \lambda_1 x_1\right) + 3(1+w_i)\Omega \right]\, , \\
   y'_2 &= \frac{y_2}{2} \left[ 6\left(x_2^2 + z^2 x_1^2\right) - \sqrt{6}\lambda_2 x_2 + 3(1+w_i)\Omega \right]\, , \\
   z'   &= \sqrt{6}\,\beta z x_2\, .
\end{align}
\end{subequations}
together with the equations for $\beta', \gamma', \lambda_i'$, \eqref{eq:betap}-\eqref{eq:lam2p}.

With this choice, however, a factor of $z^2$ appears in the denominator in $x_1'$. One possibility, as discussed in e.g.~\cite{Bahamonde:2017ize}, is to multiply the whole system of equations by $z^2$. However, this prescription does not allow to recover the fixed points in the flat axion case. Another option is to introduce $y_f = y_1/f$, as we did in the main text. However, this alternative turns out to be  also unsatisfactory: it fails to reproduce the flat axion case and does not lead to a transparent picture of the system’s dynamics. 

This exercise illustrates that apparently natural variable choices can obscure the physics of the system. Thus the $(x_f,y_f)$ parametrisation we introduced  in the main text remains the most effective framework for capturing the dynamical features of the axion--saxion system.


\addcontentsline{toc}{section}{References}
\bibliographystyle{utphys}

\bibliography{refsCDS}

\end{document}